\documentclass[11pt]{article}
\usepackage[margin=1in]{geometry}
\usepackage{amsmath,amssymb,amsthm,amsfonts,dsfont,bm}
\usepackage{graphicx}
\usepackage{subcaption}
\usepackage{longtable,booktabs,array,multirow,pdflscape,rotating}
\usepackage[plain,noend]{algorithm2e}
\usepackage{dashbox}
\usepackage[authoryear,round]{natbib}
\usepackage[colorlinks=true,linkcolor=blue,citecolor=blue,urlcolor=blue]{hyperref}
\newcommand{\I}{\mathds{1}}

\def\he{\hspace{1ex}}
\def\hee{\hspace{2ex}}
\def\eps{\varepsilon}
\makeatletter
\@ifundefined{theorem}{\newtheorem{theorem}{Theorem}}{}
\@ifundefined{lemma}{\newtheorem{lemma}{Lemma}}{}
\@ifundefined{corollary}{}{}
\@ifundefined{proposition}{}{}
\@ifundefined{definition}{}{}
\@ifundefined{remark}{}{}
\@ifundefined{keywords}{\newenvironment{keywords}{\par\medskip\noindent\textbf{Key words:} }{\par}}{}
\providecommand{\jname}[1]{}\providecommand{\jyear}[1]{}\providecommand{\jvol}[1]{}
\providecommand{\jnum}[1]{}\providecommand{\cyear}[1]{}\providecommand{\accessdate}[1]{}
\providecommand{\received}[1]{}\providecommand{\revised}[1]{}
\providecommand{\affil}[1]{}\providecommand{\email}[1]{}

\renewcommand{\markboth}[2]{}
\def\@title{Pseudo-Likelihood Regression for the Marginal Rate of Recurrent Events with a Competing Terminal Event}
\def\@author{A. Bellach\thanks{Department of Biostatistics, University of North Carolina at Chapel Hill, NC 27599, U.S.A. E-mail: abellach@unc.edu}
  \and M. R. Kosorok\thanks{Department of Biostatistics and Department of Statistics and Operations Research, University of North Carolina at Chapel Hill, NC 27599, U.S.A. E-mail: kosorok@bios.unc.edu}}
\def\@date{}
\renewcommand{\title}[2][]{}
\renewcommand{\author}[2][]{}
\renewcommand{\date}[1]{}
\makeatother

\begin{document}

\jname{Biometrika}
\markboth{A. Bellach and M. R. Kosorok}{Pseudo-Likelihood Regression for the Marginal Rate of Recurrent and Terminal Events}

\title{Pseudo-Likelihood Regression for the Marginal Rate of Recurrent Events with a Competing Terminal Event}

\author{A. BELLACH}
\affil{Department of Biostatistics, University of North Carolina at Chapel Hill,\\ Chapel Hill, NC 27599, U.S.A.
\email{abellach@unc.edu}}

\author{M. R. KOSOROK}
\affil{Department of Biostatistics and Department of Statistics and Operations Research, University of North Carolina at Chapel Hill,\\ Chapel Hill, NC 27599, U.S.A.
\email{kosorok@bios.unc.edu}}

\maketitle

\begin{abstract}
\noindent Regression modeling of recurrent and terminal events continues to present methodological challenges in survival analysis. Existing approaches either make unverifiable assumptions about the dependence between the two event types, rely on a proportional mean rate assumption, or use ad hoc estimating equations. We propose a semiparametric regression model based on a novel weighted pseudo-likelihood function that directly targets the marginal rate of the recurrent event and extends the Ghosh--Lin model to a large class of transformation models.
Our framework accommodates external time-dependent covariate effects on the marginal rate and enables flexible prediction of the marginal mean.
We establish consistency and asymptotic normality of the estimators and propose a robust sandwich estimator for the variance. In simulation studies, we demonstrate a strong performance of the proposed estimator under independent right-censoring. We apply the method to data from the interim analysis of a randomized clinical trial on chronic obstructive pulmonary disease to predict the expected number of severe exacerbations, and illustrate how the choice of regression model affects
the estimated covariate effects and predictions.
\end{abstract}


\section{Introduction}\label{sec:intro}
Recurrent events arise in clinical studies when subjects experience an event of interest multiple times. Additional challenges occur when they are also exposed to a competing terminal event. The resulting complexity is often not sufficiently accounted for in the statistical analysis. Simplified approaches either disregard subsequent events entirely or rely on count-based methods, such as negative binomial regression, which do not adequately account for censoring, terminal events, or the timing of recurrent events. Other approaches treat the terminal event as a censoring event, thereby targeting the cause-specific rate conditional on survival. When the resulting estimates are interpreted in terms of the marginal mean, it is implicitly assumed that subjects continue to experience recurrences after the terminal event, which leads to an overestimation of the marginal mean.
For example, earlier studies concluded that cardiac resynchronization therapy may increase heart failure hospitalizations. \citet{Anand:2009} showed that this conclusion was reversed when adequately accounting for recurrent and competing terminal events in the COMPANION trial using the Ghosh--Lin model \citep{GhoshLin:2002}.

\begin{figure}[t!]
	\centering
	\includegraphics[width=0.85\textwidth]{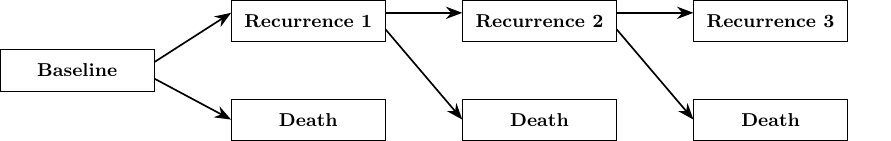}
	\caption{\small Recurrent events as a sequence of competing risk models}
	\label{fig:1}
\end{figure}

The setting with recurrent and terminal events can be interpreted as a sequence of competing risks models (Figure \ref{fig:1}), and thus poses challenges in estimation and interpretation analogous to those in the competing risks setting. Early work on competing risks focused on latent failure time models, with a major drawback that the dependence structure among competing risks needs to be specified. This dependence structure is not identifiable from observed data \citep{Slud.etal:1988}.

Contemporary approaches for competing risks data model either the cause-specific hazards or the subdistribution hazard, where the dependence structure between competing events is no longer required. For modeling the cause-specific hazards, competing risk events are removed from the risk set in the same way as censorings. Parameter estimates targeting the cause-specific hazards are not directly interpretable with regard to the cumulative incidence function, a complex function of all cause-specific hazards. In contrast, for modeling the subdistribution hazard, competing risk events are retained in the risk set, analogous to a cure fraction of subjects who will never experience the event of interest \citep{FineGray:1999, Bellach.etal:2019, Bellach.etal:2020}. As a result, the subdistribution hazard directly corresponds to the cumulative incidence function, and parameter estimates are directly interpretable.

For recurrent and terminal events, the situation is similar. Frailty models, e.g. \citep{WJC:2001, HW:2004, ZengLin:2007}, require modeling the dependence structure between event types via a frailty term. Marginal models, in contrast, allow covariate effect estimation for both endpoints without specifying their dependence structure. \citet{CookLaw:2018} established semiparametric regression modeling for the cause-specific rate conditional on survival, a framework also discussed by \citet{GhoshLin:2002}. However, regression coefficients do not have a direct interpretation with regard to the expected number of recurrences, because of the complex, nonlinear relation between the targeted conditional rate and the marginal mean, and prediction would require correctly specified models for both the recurrent and terminal events.

\citet{AAR:2019} adapt pseudo-value techniques to model marginal features. Their framework relies on applying generalized estimating equations to the constructed pseudo-values. Consequently, the flexible approach depends on a pre-specified time grid and lacks a continuous-time regression framework for the expected number of events.

\citet{GhoshLin:2000} propose nonparametric estimation of the marginal rate function, which has a direct correspondence with the marginal mean of the recurrent events. Semiparametric regression modeling is established in \cite{GhoshLin:2002} based on the key idea that subjects with a previous terminal event are incorporated as a cure fraction into the risk set as in \cite{FineGray:1999}. The Ghosh--Lin model therefore is the natural extension of the Fine--Gray model for the subdistribution of a competing risk to scenarios with recurrent and terminal events. Their approach is based on a weighted estimating equation and imposes a proportional rates assumption.

Our proposed pseudo-likelihood approach allows for flexible regression modeling within a broad class of semiparametric transformation models and extends the Ghosh--Lin model for the marginal rate. Parameter estimates are interpretable with regard to the marginal mean and individualized prediction is straightforward. In the presence of terminal events, the treatment effect typically changes over time, resulting in a violation of the proportionality assumption imposed by the Ghosh--Lin model and a flatter trajectory of the expected marginal mean. These deviations are directly addressed by our flexible regression model.    

By incorporating subjects with a previous terminal event as a cure fraction into the pseudo-risk set, our approach reweights the contribution of each recurrent event and does not require specifying the dependence structure between recurrent and terminal events. Because the estimator maximizes a pseudo-likelihood function, the Takeuchi information criterion is immediately available for model selection \citep{Takeuchi:1976, KonishiKitagawa:1996}. Our framework further accommodates external time-dependent covariates, extending the Ghosh--Lin approach designed for time-fixed covariates.

In Section~\ref{sec:wlikely}, we construct the pseudo-likelihood from
pseudo-subjects with left-truncated entry times, which partly accounts
for within-subject dependence. Residual dependence does not affect
consistency and is accounted for by a sandwich estimator for the
variance, proposed in Section~\ref{sec:asymprop}. In Appendix~\ref{Appendix}, 
we prove uniform consistency and weak convergence to a
Gaussian process for the Box--Cox and logarithmic transformation models using
empirical process techniques that do not rely on traditional martingale
structures. In Section~\ref{sec:ghosh-lin}, we derive the ad hoc estimating equation proposed by Ghosh and Lin through factorization of our proposed pseudo-likelihood function under the identity transformation and with time-fixed covariates. Consequently, both frameworks target the same population-level marginal mean parameters in this setting. 

\section{Semiparametric Transformation Models for the Marginal Mean}
\label{sec:semitrans}
Let $N^*(t)$ denote the process counting the number of recurrences in $[0,t]$, and let $D$ denote the terminal event time. Given a time-fixed covariate vector $Z\in\mathbb{R}^d$, the marginal mean $E\{N^*(t)\mid Z\}$ can be derived via an Aalen--Johansen type estimator \citep{CookLaw:1997, GhoshLin:2000}. This approach relies on the cause-specific rate conditional on survival, $\lambda^*(s\mid Z)=E\{dN^*(s)\mid D\geq s,Z\}$, and the survival function for the terminal event, $S_D(t\mid Z)=P(D>t\mid Z)$. In this cause-specific framework, subjects experiencing the terminal event are removed from the risk set in the same way as censored subjects, thereby restricting the interpretation of \(\lambda ^{*}\) to the surviving cohort. Prediction of the marginal mean then requires a plug-in estimator that is sensitive to misspecification of the models for either $\lambda^*$ or $S_D$. 

Our proposed regression model is based on a pseudo-likelihood function and specified via the marginal rate function, proposed by \citet{GhoshLin:2000,GhoshLin:2002},
\begin{equation*}
	\lambda\{t\mid Z(t)\}=E\{dN^*(t)\mid Z(u),\,0< u\leq t\}.
\end{equation*}
The marginal rate does not depend on the prior event history, and the model therefore
makes no assumption about within-subject dependencies. Because $\lambda$ is defined without conditioning on survival, it has a direct correspondence with the marginal mean with external covariates,
\[
E\{N^*(t)\mid Z(u), 0<u\leq t\}=\int_0^t\lambda\{u\mid Z(u)\}\,du.
\]
In Section \ref{sec:wlikely} we establish that $\lambda$ is associated with a pseudo-risk set that reweights the contribution of recurrent events with regard to previous terminal events. As a result of the reweighting, the terminal events are already incorporated in the marginal mean estimators and a specification of the dependence structure between recurrent and terminal events is not required.

Let $\beta\in\mathbb R^d$ denote a vector of unknown regression parameters, $\Lambda(t)$ an unspecified increasing cumulative baseline rate, and $\mathcal G$ a thrice continuously differentiable and strictly increasing link function with $\mathcal G(0)=0$, $\mathcal G'(0)>0$ and $\mathcal G(\infty)=\infty$. We propose the general model for the cumulative marginal rate 
\begin{equation}\label{eq:MMcum}
	\Lambda\left\{t\mid Z(t)\right\}=\mathcal G\left\{\int_0^t e^{\beta^T Z(s)}d\Lambda(s)\right\},
\end{equation}
introduced by \citet{ZengLin:2006} for recurrent event models without competing terminal events and further explored by \citet{Bellach.etal:2019,Bellach.etal:2020} for the subdistribution hazard of a competing risk. In Section \ref{sec:Acondition} we specify additional regularity conditions, which are considerably weaker than those in \citet{ZengLin:2006}. Special cases include the Box--Cox transformation models where $\mathcal G(x)=\left\{(1+x)^\rho-1\right\}/\rho$ with $\rho\geq 0$, and the logarithmic transformation models where $\mathcal G(x)=\log(1+rx)/r$ with $r\geq 0$ \citep{Chen.etal:2002}. The proportional rate model is the Box--Cox transformation model with $\rho=1$ and the logarithmic transformation model as $r\to 0$. The proportional odds model is the logarithmic transformation model with $r=1$ and the Box--Cox transformation model as $\rho\to 0$. The marginal rate function
\begin{equation}\label{eq:MMinst}
	\lambda\left\{t\mid Z(t)\right\}=e^{\beta^TZ(t)}\lambda(t)\mathcal G'\left\{\int_0^t e^{\beta^TZ(s)}d\Lambda(s)\right\},
\end{equation}
with $\lambda(t)=d\Lambda(t)/dt$ denoting the baseline marginal rate.

\section{Pseudo-Likelihood Function for the Marginal Rate}
\label{sec:wlikely}
We derive a pseudo-likelihood function for the marginal rate in
which the contribution of a subject with $k$ recurrences and one terminal 
event or censoring is written as the contributions of $k+1$ pseudo-subjects with 
left-truncated entry times. Within-subject dependence is thereby partly 
accounted for through the left-truncated entry times and the shared covariates. 
The working assumption is that, given the time of the previous recurrent event 
and the shared covariates, the rate of a new recurrence does not otherwise depend 
on the subject's history. Any residual dependence affects only the variability 
of the estimators, not their consistency, which holds whenever the marginal rate model is correctly specified, as in \citet{LawlessNadeau:1995}, \citet{Linetal:2000} 
and \citet{WellnerZhang:2007}. The sandwich variance estimator proposed in Section~\ref{sec:asymprop} accounts for this residual dependence and for the additional variability from the estimated censoring weights.

The terminal event may be an absorbing event or a cure event. The
framework for recurrent and cure events under independent
right-censoring is mathematically equivalent to that for recurrent and
absorbing events under administrative right-censoring. For modeling the
marginal rate, subjects with a previous terminal event are
retained in the risk set, which is intuitively appealing for cure
events. For absorbing events, the key idea is to mimic the cure-event
setting. The difference is that right-censoring can be observed after a
cure event but not after an absorbing event. To obtain the pseudo-risk
set for recurrent and absorbing events, we apply inverse
probability of censoring weighting to account for the unobserved
right-censoring of subjects in the cure fraction.

Formally, for an iid sample of $n$ subjects, let $T_{ij}$ $(i=1,\ldots,n;\,j=1,\ldots,n_i)$ denote the observed event times, and let $\varepsilon_{ij} \in \{1,2\}$ indicate the event type, with
$\varepsilon_{ij}=1$ denoting a recurrence and $\varepsilon_{ij}=2$ the
terminal event. Let $N^*(t)=\sum_{i=1}^n N_i^*(t)$ represent the process counting the total recurrent events, where $N_i^*(t)=\sum_{j=1}^{n_i}\mathbf{1}(T_{ij}\leq t, \varepsilon_{ij}=1)$. The study endpoint is denoted by $\tau$. 
\subsection{Recurrent and Terminal Events with Complete Follow-up}\label{sec:complete}
Modeling the marginal rate, subjects with a previous terminal event remain in the risk set until $\tau$. The subject-level at-risk indicator is thus defined for both scenarios as $Y_i(t)=1(D>t)+1(D\leq t)=1$\, for \,$t\leq\tau$, $i=1,\ldots,n$, and $Y(t)=\sum_{i=1}^n Y_i(t)=n$ is the size of the risk set $\mathcal R(t)=\{i:\,\tau\geq t\}$. 
Then $E\{dN_i^*(t)\} = \lambda(t)Y_i(t)\,dt$, and from the decomposition
$dN_i^*(t) = \lambda(t)Y_i(t)\,dt + dM_i(t)$, where $M_i$ has mean zero,
we obtain the Nelson--Aalen-type estimator for the cumulative marginal
rate, $\widehat\Lambda(t) := n^{-1}\int_0^t dN^*(u)$. The process $M_i$ has
mean zero but is in general not a martingale with respect to the
subject's own history, because the rate of a new recurrence may depend
on previous recurrences beyond the marginal rate. Nevertheless,
the asymptotic properties can be derived with empirical process
techniques for iid subject-level processes \citep{VanderVaart.etal:2023, Kosorok:2008}, as in
\citet{Linetal:2000}.

Developing the likelihood, we build on the idea that subjects re-enter the likelihood after each recurrent event. For a subject with $n_i$ recurrences and administrative censoring at $\tau$, the likelihood contribution is
\begin{align*}
	&\lambda(T_{i1})S(T_{i1})
	\cdot \lambda(T_{i2})\bigl\{S(T_{i2})/S(T_{i1})\bigr\}
	\cdot\ldots\cdot
	\lambda(T_{in_i})\bigl\{S(T_{in_i})/S(T_{in_{i-1}})\bigr\} \cdot\bigl\{S(\tau)/S(T_{in_i})\bigr\}
\end{align*}
with $S(t)=\exp\{-\Lambda(t)\}$, so that under the working assumption
$S(t)$ and $S(t)/S(s) = \exp[-\{\Lambda(t)-\Lambda(s)\}]$ are the
probabilities that no recurrent event has occurred until $t$ and in
$(s,t]$, respectively. For a subject with $n_i-1$ recurrences and a
terminal event at $T_{in_i}$ the contribution is
\begin{align*}
	&\lambda(T_{i1})S(T_{i1})
	\cdot\ldots\cdot
	\lambda(T_{in_{i-1}})\bigl\{S(T_{in_{i-1}})/S(T_{in_{i-2}})\bigr\} \cdot\bigl\{S(T_{in_i})/S(T_{in_{i-1}})\bigr\}\cdot\bigl\{S(\tau)/S(T_{in_i})\bigr\}.
\end{align*}
With $\Delta_{ij}=\mathbf{1}(T_{ij}\leq\tau)$, the likelihood function then takes the form
\[
	L_n=\prod_{i=1}^n S(\tau) \,\prod_{j=1}^{n_i}\,
	\lambda(T_{ij})^{\Delta_{ij} \I\{\varepsilon_{ij}=1\}}
	=\prod_{i=1}^n \exp\bigl\{-\Lambda(\tau)\bigr\}\,\prod_{j=1}^{n_i}\,\lambda(T_{ij})^{\Delta_{ij} \I\{\varepsilon_{ij}=1\}} ,
\]
regardless of whether the terminal events are absorbing or cure events. 
\subsection{Recurrent and Cure Events with Independent Right-Censoring}\label{sec:curewcens}
Let $C\leq\tau$ denote the censoring time, $X=T\wedge D\wedge C$ and $\Delta=\I\{T\leq C\wedge\tau\}$. For an iid sample of subjects $i=1,\ldots,n$ with recurrences $j=1,\ldots,n_i$, let $C_i$ denote the individual censoring time, $X_{ij}=T_{ij}\wedge C_i\wedge D_i$, and $\Delta_{ij}=\I\{T_{ij}\leq C_{i}\wedge\tau\}$. Let $N(t)=\sum_{i=1}^n N_i(t)$ denote the process counting the recurrent events, where $N_i(t)=\int_0^t \I\{C_i\geq u\}dN_i^*(u)$ for $i=1,\ldots,n$. Let $Y_i(t)=\I\{C_i\wedge\tau\geq t\}$ denote the at-risk indicator for $i=1,\ldots,n$, then $Y(t)=\sum_{i=1}^n Y_i(t)$ is the size of the risk set $\mathcal R(t)=\{i:\,C_i\wedge\tau\geq t\}$. Under independent right-censoring, $E\{dN_i(t)\} = \lambda(t)Y_i(t)\,dt$, and the decomposition $dN_i(t) = \lambda(t)Y_i(t)\,dt + dM_i(t)$ 
holds with $M_i$ of mean zero, as in Section~\ref{sec:complete}. 
With similar arguments as in Section~\ref{sec:complete}, the likelihood function for the marginal rate takes the form
\[
L_n=\prod_{i=1}^n \exp\bigl\{- \Lambda(C_i\wedge\tau)\bigr\}\,\prod_{j=1}^{n_i}\,\lambda(X_{ij})^{\Delta_{ij} \I\{\varepsilon_{ij}=1\}}.
\]
\subsection{Recurrent and Absorbing Events with Independent Right-Censoring}\label{recabs}
To mimic the structure of the likelihood in Section \ref{sec:curewcens}, we incorporate subjects with a previous absorbing event as a cure fraction into the pseudo-risk set. To account for unobserved right-censoring after the absorbing event, we apply the weight function $w_i(t)=\I(C_i>D_i\wedge t)G_c(t)/G_c(D_i\wedge t)$, where in practice the censoring distribution $G_c$ is replaced by its Kaplan--Meier estimator $\widehat{G}_c$, as proposed by \cite{GhoshLin:2002}. A Nelson--Aalen type estimator for the marginal rate can be derived from the weighted decomposition $w_i(t)dN_i(t)=w_i(t)Y_i(t)\lambda(t)dt+w_i(t)dM_i(t)$ with
\begin{equation}
	\sum_{i=1}^n w_i(t)Y_i(t)=\sum_{i=1}^n \I\{D_i\wedge C_i\geq t\} + \sum_{i=1}^n\I\{D_i\leq t\}
	\cdot\bigl\{G_c(t)\slash G_c(D_i)\bigr\}
\end{equation}
representing the expected number of subjects in the pseudo-risk set $\mathcal R_p(t)$.
Applying these weights, the pseudo-likelihood function takes the form
\[
	L_n =\prod_{i=1}^n\exp\Bigl\{-\int_0^\tau w_i(u)Y_i(u)d\Lambda(u)\Bigr\}\cdot\prod_{j=1}^{n_i}\lambda(X_{ij})^{\Delta_{ij} \I\{\varepsilon_{ij}=1\}}.
\]
The Nelson--Aalen estimator for the marginal rate can be derived from the product-integral representation of the likelihood function in Sections \ref{sec:complete},\,\ref{sec:curewcens}, and \ref{recabs}. Utilizing nonparametric maximum likelihood techniques, $\Lambda$ is therefore approximated by a step function with point masses at the observed recurrent event times, and the nonparametric pseudo-likelihood function is maximized with respect to jump sizes $\Lambda\{t\} := \Lambda(t) - \Lambda(t-)$. 
\section{Semiparametric Regression Model}
Having established the weighted pseudo-likelihood function corresponding to the marginal rate, we now incorporate the general models \eqref{eq:MMcum} and \eqref{eq:MMinst} to derive the log-likelihood function for the semiparametric regression model.
\subsection{Nonparametric Pseudo-Likelihood Estimation}
For the recurrent event model with competing cure events and independent right-censoring the log-likelihood function takes the form
\begin{eqnarray*}
	\ell(\beta,\Lambda)
	&=&\sum_{i=1}^n\left(\int_0^\tau \log\left[Y_i(t)e^{\beta^TZ_i(t)}\lambda(t)\,
	\mathcal G'\left\{\int_0^t e^{\beta^TZ_i(u)}d\Lambda(u)\right\}\right]\I(C_i\geq t)dN_i^*(t)\right.\\
	&&\hspace{1.5cm}-\left.\int_0^\tau \I(C_i\geq t)Y_i(t)e^{\beta^TZ_i(t)}
	\mathcal G'\left\{\int_0^t e^{\beta^TZ_i(u)}d\Lambda(u)\right\}d\Lambda(t)\right).
\end{eqnarray*}
For the model with competing terminal events and independent right-censoring we obtain
\begin{eqnarray*}
	\ell(\beta,\Lambda)
	&=&\sum_{i=1}^n\left(\int_0^\tau \log\left[Y_i(t)e^{\beta^TZ_i(t)}\lambda(t)\,
	\mathcal G'\left\{\int_0^t e^{\beta^TZ_i(u)}d\Lambda(u)\right\}\right]\I(C_i\geq t)dN_i^*(t)\right.\\[1ex]
	&&\hspace{1.5cm}-\left.\int_0^\tau w_i(t)Y_i(t)e^{\beta^TZ_i(t)}
	\mathcal G'\left\{\int_0^t e^{\beta^TZ_i(u)}d\Lambda(u)\right\}d\Lambda(t)\right).
\end{eqnarray*}
\normalsize
\noindent
Decomposing the pseudo-risk set as $w_i(t)Y_i(t)=\I(D_i\geq t)+w_i^*(t)\I(D_i\leq t)$ with a simplified weight function $w_i^*(t)=\widehat{G}_c(t)/\widehat{G}_c(D_i)$ we obtain the representation  
\begin{eqnarray}
	\ell(\beta, \Lambda)&=&\hspace{-0.8ex}\sum_{i=1}^n\left(\int_0^\tau \log\left[Y_i(t)e^{\beta^TZ_i(t)}\lambda(t)\,
	\mathcal G'\left\{\int_0^t e^{\beta^TZ_i(u)}d\Lambda(u)\right\}\right]\I(C_i\geq t)dN_i^*(t)\right)\nonumber\\
	&&-\hee\sum_{i=1}^n\he
	\mathcal G\left\{\int_0^{D_i\wedge C_i\wedge\tau}e^{\beta^TZ_i(t)}d\Lambda(t)\right\}
	\nonumber\\
	&&-\sum_{i:D_i<\tau}\left[\int_{D_i}^\tau w_i^*(t)e^{\beta^TZ_i(t)}
	\mathcal G'\left\{\int_0^t e^{\beta^TZ_i(u)}d\Lambda(u)\right\}d\Lambda(t)\right].\label{likenew}
\end{eqnarray}
To maximize the pseudo-likelihood nonparametrically, $\Lambda$ is approximated by a step function with jumps at the observed recurrences (Section~\ref{sec:calc}). The estimator $\widehat{\theta}_n=(\widehat{\beta}_n,\widehat{\Lambda}_n)$ is then obtained by maximizing the log-pseudo-likelihood with respect to $\beta$ and the jump sizes $\Lambda\{X_{ij}\}$, using automatic differentiation in \texttt{R} \citep{Kristetal:2016, Rteam}.

\subsection{The Ghosh--Lin estimating equation as a special case}
\label{sec:ghosh-lin} 
Under the identity transformation and with time-fixed covariates, the
marginal rate reduces to $\lambda(t \mid Z) = e^{\beta^T Z}\lambda(t)$
and the weighted pseudo-likelihood function can be decomposed as
\begin{eqnarray}
	L_n &=&\notag \prod_{i=1}^n\Biggl[\prod_{s\leq\tau}\bigl\{Y_i(s)e^{\beta^TZ_i(s)}\lambda(s)\bigr\}^{dN_i^*(s)}
	\cdot\exp\biggl\{-\int_0^\tau w_i(u)Y_i(u)e^{\beta^TZ_i(u)}\,d\Lambda(u)\biggr\}\Biggr]\\[1ex]
	&=&\notag\prod_{i=1}^n\Biggl(\prod_{s\leq\tau}
	\Biggl[\frac{w_i(s)Y_i(s)e^{\beta^TZ_i(s)}}
	{\sum_{j=1}^n w_j(s)Y_j(s)e^{\beta^TZ_j(s)}} 
	\cdot\sum_{j=1}^n\bigl\{w_j(s)Y_j(s)e^{\beta^TZ_j(s)}\lambda(s)\bigr\}\Biggr]^{dN_i^*(s)}\\[1ex]
	&&\hspace{4cm}\quad\times\exp\biggl\{-\int_0^\tau w_i(u)Y_i(u)e^{\beta^TZ_i(u)}\,d\Lambda(u)\biggr\}\Biggr).
\end{eqnarray}
This factorization generalizes the classical likelihood decomposition by \citet{Efron:1977}.
The first term is a weighted partial likelihood function. It has a direct interpretation as a product of the conditional probabilities that a specific subject has a recurrent event at time $s$ given that any subject in the pseudo-risk set $\mathcal R_p$ has a recurrent event at $s$. The corresponding weighted score equation is equivalent to the weighted estimating equation proposed by \citet{GhoshLin:2002},
\[
	U(\beta)=\sum_{i=1}^n\int_0^\tau\left[Z_i(s)-\frac{\sum_j w_j(s)Y_j(s)Z_j(s)\exp\{Z_j(s)\beta\}}{\sum_j w_j(s)Y_j(s)\exp\{Z_j(s)\beta\}}\right]\,w_i(s)dN_i^*(s),
\]
For recurrent and cure events the score equation takes the form
\[
	U^c(\beta)=\sum_{i=1}^n\int_0^\tau\left[Z_i(s)-\frac{\sum_j\I(C_j\geq s)Y_j(s)Z_j(s)\exp\{Z_j(s)\beta\}}{\sum_j\I(C_j\geq s)Y_j(s)\exp\{Z_j(s)\beta\}}\right]\, \I\{C_i\geq s\}dN_i^*(s).
\]
\subsection{Asymptotic Properties and Variance Estimation}
\label{sec:asymprop}
In the Appendix and Supplementary Material we prove the following
\begin{theorem}\label{theorem:1} The estimator $\widehat{\theta}_n=(\widehat{\beta}_n,\widehat{\Lambda}_n)$ derived from
	maximizing the weighted pseudo-likelihood function is uniformly consistent.
\end{theorem}
Using Lemma~\ref{lemma3} from Section~\ref{sec:lemma3}, which extends Theorem 3.3.1 by \citet{VanderVaart.etal:2023} to weighted $\mathcal{Z}$-estimators, we prove
\begin{theorem}\label{theorem:2} $\sqrt{n}\bigl(\widehat{\beta}_n-\beta_0,\widehat{\Lambda}_n-\Lambda_0\bigr)$
converges weakly to a Gaussian process.
\end{theorem}
We extend the theory developed in \citet{Bellach.etal:2019, Bellach.etal:2020} for the subdistribution of a competing risk to settings with recurrent and terminal events, adapting  key arguments of \citet{Murphy.etal:1994, Murphy.etal:1995}, \citet{Parner:1998}, and \citet{ZengLin:2006}. For weak convergence and variance estimation we evaluate linear functionals
\[
\sqrt{n}(\widehat{\theta}_n-\theta_0)(h)=\sqrt{n}\,h_1^T(\widehat{\beta}_n-\beta_0)+\sqrt{n}\int_0^\tau h_2(t)d\{\widehat{\Lambda}_n(t)-\Lambda_0(t)\}, 
\]
where $h=(h_1,h_2)$, $h_2$ belongs to the Skorohod space $\mathcal D[0,\tau]$ and $h_1\in\mathbb R^d$. Let $\{0<\widetilde{T}_1<\widetilde{T}_2<\ldots<\widetilde{T}_{k}\}$ denote the ordered recurrent event times and $k$ the number of observed recurrent events. We define a corresponding vector $\widetilde{h}_2\in\mathbb{R}^k$ with $\widetilde{h}_2^i=h_2(\widetilde{T}_i)$ and $\widetilde{\Lambda}\{\widetilde{T}_i\}=\Lambda\{\widetilde{T}_i\}-\Lambda\{\widetilde{T}_{i-1}\}$ denote the jump sizes for $i=1,\ldots,k$. With $\widetilde{h}=(h_1,\widetilde{h}_2^1,\ldots,
\widetilde{h}_2^k)\in\mathbb{R}^{d+k}$ the asymptotic variance of the linear functional can be estimated by the sandwich estimator 
\begin{equation}
	\widehat{\mbox{Var}}\sqrt{n}(\widehat{\theta}_n-\theta_0)(\widetilde{h})
	=\widetilde{h}^T\,\bigl(\mathcal I_n^{-1}\bigr)^T\,\widehat{\Sigma}_n\,\,\mathcal I_n^{-1}\widetilde{h},
\end{equation}
where $\mathcal I_n$ is the observed Fisher information with respect
to $\beta$ and jump sizes, and
$\widehat{\Sigma}_n=n^{-1}\sum_{i=1}^n(\hat{\eta}_i+\hat{\kappa}_i)^{\otimes 2}$ is the estimated variance of the score with respect to $\beta$ and jump sizes, with $\hat{\eta}_i$ and $\hat{\kappa}_i$ denoting the components of the iid decomposition provided in Section \ref{sec:varest}. Analogously we obtain the
estimator for the covariances of $\sqrt{n}(\widehat{\theta}_n-\theta_0)(h)$, $\sqrt{n}(\widehat{\theta}_n-\theta_0)(g)$, with $h,g\in\mathbb{R}^d\times \mathcal D[0,\tau]$
and with corresponding vectors $\widetilde{g},\widetilde{h}\in\mathbb{R}^{d+k}$,
\begin{equation}
	\widehat{\mbox{Cov}}\big\{\sqrt{n}(\widehat{\theta}_n-\theta_0)(\widetilde{h}),\sqrt{n}(\widehat{\theta}_n-\theta_0)(\widetilde{g})\big\}=\widetilde{g}^T\,\bigl(\mathcal I_n^{-1}\bigr)^T\,\widehat{\Sigma}_n\,\,\mathcal I_n^{-1}\widetilde{h}. 
\end{equation}
For model selection under independent right-censoring, we adapt the Takeuchi information criterion \citep{Takeuchi:1976, KonishiKitagawa:1996}, with the penalty term computed from the components of the sandwich estimator,
\begin{equation}
	\mathrm{TIC} = -2\,\ell_n(\hat\theta_n) + 2\,\mathrm{tr}\big(\mathcal{I}_n^{-1}\hat\Sigma_n\big),
\end{equation}
where $\ell_n$ is the empirical version of the log-pseudo-likelihood, defined in Section~\ref{sec:calc}.
For recurrent and cure events, or absorbing events with administrative censoring, no estimated weights are required, and the sandwich estimator only needs to account for the residual dependence among a subject's recurrences. In many settings, the Akaike information criterion \citep{Akaike:1974} provides a good approximation of the Takeuchi information criterion, see Section~\ref{sec:sims}.

\section{Simulation studies}\label{sec:sims}
We conducted simulation studies for the Box--Cox and logarithmic classes of transformation models. To the best of our knowledge, this is the first simulation study with data generated directly from the marginal rate formulation, while \citet{GhoshLin:2002} and others have simulated such data from a frailty framework. Censoring times were generated from a uniform distribution, with parameters specified for each simulation scenario in the Supplementary Material. Covariates $Z_{i1}$ and $Z_{i2}$ were drawn independently from a standard normal distribution for \(i=1,\ldots,n\).
%
\begin{landscape}
	\begin{table}[p]
		\centering
		\footnotesize
		\setlength{\tabcolsep}{4pt}
		\renewcommand{\arraystretch}{1.05}
		\caption{\bf Results from simulation studies for the Box--Cox transformation models based on 10\,000 replicates}
		\label{tab:sim_prop_rho}
		\resizebox{\linewidth}{!}{%
			\begin{tabular}{llrrrrrrrrrrrrrrrrrr}
				&  & \multicolumn{6}{c}{$\rho=0.5$}
				& \multicolumn{6}{c}{$\rho=1$}
				& \multicolumn{6}{c}{$\rho=2$} \\

				$n$ & Par.
				& Bias(\%) & SD & SE$_F$ & SE$_S$ & CP$_F$ & CP$_S$
				& Bias(\%) & SD & SE$_F$ & SE$_S$ & CP$_F$ & CP$_S$
				& Bias(\%) & SD & SE$_F$ & SE$_S$ & CP$_F$ & CP$_S$ \\[4pt]

				50
				& $\beta_1$
				& 0.007\he(0.7) & 0.156 & 0.153 & 0.140 & 0.949 & 0.911
				& -0.003\he(0.3) & 0.130 & 0.127 & 0.115 & 0.952 & 0.912
				& 0.001\he(0.1) & 0.101 & 0.098 & 0.090 & 0.947 & 0.913 \\

				& $\beta_2$
				& 0.000\he(0.2) & 0.148 & 0.146 & 0.135 & 0.950 & 0.912
				& 0.001\he(0.3) & 0.121 & 0.117 & 0.106 & 0.944 & 0.902
				& 0.001\he(0.3) & 0.074 & 0.072 & 0.063 & 0.953 & 0.890 \\

				& $A(\tau\!/\!4)$
				& -0.012\he(1.2) & 0.187 & 0.185 & 0.180 & 0.933 & 0.921
				& -0.001\he(0.2) & 0.084 & 0.083 & 0.080 & 0.939 & 0.927
				& -0.001\he(0.2) & 0.057 & 0.057 & 0.055 & 0.943 & 0.929 \\

				& $A(\tau\!/\!2)$
				& -0.012\he(0.7) & 0.272 & 0.268 & 0.260 & 0.937 & 0.925
				& 0.000\he(0.0) & 0.124 & 0.122 & 0.119 & 0.940 & 0.929
				& 0.000\he(0.0) & 0.076 & 0.076 & 0.074 & 0.945 & 0.931 \\

				& $A(\tau)$
				& 0.010\he(0.4) & 0.354 & 0.350 & 0.340 & 0.942 & 0.931
				& 0.009\he(0.8) & 0.179 & 0.175 & 0.171 & 0.944 & 0.931
				& 0.003\he(0.4) & 0.097 & 0.096 & 0.094 & 0.941 & 0.929 \\

				\addlinespace

				100
				& $\beta_1$
				& 0.004\he(0.4) & 0.105 & 0.105 & 0.100 & 0.952 & 0.932
				& -0.004\he(0.4) & 0.087 & 0.084 & 0.079 & 0.946 & 0.923
				& -0.002\he(0.2) & 0.066 & 0.065 & 0.061 & 0.946 & 0.924 \\

				& $\beta_2$
				& -0.002\he(0.8) & 0.101 & 0.100 & 0.096 & 0.948 & 0.933
				& 0.000\he(0.0) & 0.080 & 0.078 & 0.074 & 0.948 & 0.921
				& 0.000\he(0.1) & 0.048 & 0.047 & 0.043 & 0.952 & 0.914 \\

				& $A(\tau\!/\!4)$
				& -0.005\he(0.5) & 0.131 & 0.131 & 0.129 & 0.945 & 0.941
				& 0.000\he(0.0) & 0.059 & 0.058 & 0.057 & 0.942 & 0.935
				& 0.000\he(0.1) & 0.041 & 0.040 & 0.039 & 0.943 & 0.935 \\

				& $A(\tau\!/\!2)$
				& 0.000\he(0.0) & 0.191 & 0.189 & 0.187 & 0.943 & 0.939
				& 0.001\he(0.2) & 0.086 & 0.086 & 0.085 & 0.945 & 0.939
				& 0.001\he(0.3) & 0.054 & 0.053 & 0.052 & 0.947 & 0.939 \\

				& $A(\tau)$
				& 0.022\he(1.0) & 0.250 & 0.247 & 0.244 & 0.946 & 0.942
				& 0.011\he(1.0) & 0.125 & 0.123 & 0.122 & 0.944 & 0.940
				& 0.005\he(0.7) & 0.068 & 0.067 & 0.066 & 0.943 & 0.938 \\

				\addlinespace

				200
				& $\beta_1$
				& 0.002\he(0.2) & 0.073 & 0.073 & 0.071 & 0.952 & 0.941
				& -0.005\he(0.5) & 0.058 & 0.058 & 0.056 & 0.951 & 0.939
				& -0.004\he(0.4) & 0.045 & 0.045 & 0.043 & 0.948 & 0.937 \\

				& $\beta_2$
				& -0.001\he(0.6) & 0.071 & 0.070 & 0.069 & 0.950 & 0.938
				& 0.000\he(0.2) & 0.055 & 0.054 & 0.052 & 0.950 & 0.936
				& 0.001\he(0.3) & 0.033 & 0.032 & 0.030 & 0.946 & 0.922 \\

				& $A(\tau\!/\!4)$
				& -0.002\he(0.2) & 0.092 & 0.092 & 0.092 & 0.948 & 0.945
				& 0.001\he(0.3) & 0.041 & 0.041 & 0.041 & 0.947 & 0.943
				& 0.001\he(0.4) & 0.028 & 0.028 & 0.028 & 0.954 & 0.950 \\

				& $A(\tau\!/\!2)$
				& 0.003\he(0.2) & 0.134 & 0.134 & 0.133 & 0.948 & 0.946
				& 0.004\he(0.5) & 0.061 & 0.061 & 0.060 & 0.950 & 0.947
				& 0.003\he(0.6) & 0.037 & 0.038 & 0.037 & 0.949 & 0.948 \\

				& $A(\tau)$
				& 0.027\he(1.3) & 0.174 & 0.174 & 0.174 & 0.952 & 0.948
				& 0.014\he(1.2) & 0.086 & 0.087 & 0.087 & 0.951 & 0.950
				& 0.007\he(0.9) & 0.047 & 0.047 & 0.047 & 0.949 & 0.947 \\

				\addlinespace

				400
				& $\beta_1$
				& 0.001\he(0.1) & 0.052 & 0.051 & 0.050 & 0.948 & 0.942
				& -0.005\he(0.5) & 0.041 & 0.040 & 0.040 & 0.947 & 0.939
				& -0.005\he(0.5) & 0.031 & 0.031 & 0.031 & 0.946 & 0.941 \\

				& $\beta_2$
				& -0.002\he(0.8) & 0.049 & 0.049 & 0.049 & 0.950 & 0.947
				& 0.001\he(0.3) & 0.038 & 0.038 & 0.037 & 0.948 & 0.942
				& 0.001\he(0.3) & 0.022 & 0.022 & 0.021 & 0.948 & 0.936 \\

				& $A(\tau\!/\!4)$
				& -0.001\he(0.1) & 0.066 & 0.065 & 0.065 & 0.946 & 0.945
				& 0.002\he(0.6) & 0.029 & 0.029 & 0.029 & 0.954 & 0.952
				& 0.002\he(0.5) & 0.020 & 0.020 & 0.020 & 0.951 & 0.949 \\

				& $A(\tau\!/\!2)$
				& 0.006\he(0.4) & 0.096 & 0.094 & 0.094 & 0.945 & 0.944
				& 0.005\he(0.7) & 0.042 & 0.043 & 0.043 & 0.953 & 0.951
				& 0.003\he(0.6) & 0.026 & 0.026 & 0.026 & 0.952 & 0.953 \\

				& $A(\tau)$
				& 0.031\he(1.4) & 0.124 & 0.123 & 0.123 & 0.945 & 0.945
				& 0.015\he(1.3) & 0.062 & 0.061 & 0.061 & 0.946 & 0.946
				& 0.007\he(0.9) & 0.033 & 0.033 & 0.033 & 0.950 & 0.950 \\

			\end{tabular}
		}

		\vspace{2.5ex}

		\parbox{\linewidth}{\footnotesize Bias (Bias in \%), empirical standard deviation (SD), average estimated standard deviations based on inverse Fisher Information (SE$_F$) and Sandwich Estimator (SE$_S$), coverage probabilities (CP$_F$, CP$_S$). We investigated the following parameters and settings, with regression parameters $\beta = (1, -0.2)$ in all scenarios:\\[2ex]
			a) $\rho=0.5$, $\gamma_1=2.5$, $\gamma_2=0.4$, $\gamma_4=0.05$, $\tau=5$, $(A(\tau\!/4),\,A(\tau\!/2),\,A(\tau))=(0.984,1.580,2.162)$, avg. 2.0 recurrences, 12\% terminal events, 15\% censorings \\ \he b) $\rho=1$,\,\he $\gamma_1=1.8$, $\gamma_2=0.2$, $\gamma_4=0.1$, $\tau=5$,  $(A(\tau\!/4),\,A(\tau\!/2),\,A(\tau))=(0.398,0.708,1.138)$, avg. 1.8 recurrences, 14\% terminal events, 15\% censorings \\ \, c) $\rho=2$, $\gamma_1=0.9$, $\gamma_2=0.4$, $\gamma_4=0.05$, $\tau=5$, $(A(\tau\!/\!4), A(\tau\!/\!2), A(\tau))=(0.354,0.569,0.778)$, avg.\ 2.4 recurrences, 12\% terminal events, 15\% censorings}

	\end{table}
\end{landscape}

%
\begin{landscape}
	\begin{table}[p]
		\centering
		\footnotesize
		\setlength{\tabcolsep}{4pt}
		\renewcommand{\arraystretch}{1.05}
		\caption{\bf Results from simulation studies for logarithmic transformation models based on 10\,000 replicates}
		\label{tab:sim_log_r_full}
		\resizebox{\linewidth}{!}{%
			\begin{tabular}{llrrrrrrrrrrrrrrrrrr}
				&  & \multicolumn{6}{c}{$r=0.5$}
				& \multicolumn{6}{c}{$r=1$}
				& \multicolumn{6}{c}{$r=2$} \\

				$n$ & Par.
				& Bias\he(\%) & SD & SE$_F$ & SE$_S$ & CP$_F$ & CP$_S$
				& Bias\he(\%) & SD & SE$_F$ & SE$_S$ & CP$_F$ & CP$_S$
				& Bias\he(\%) & SD & SE$_F$ & SE$_S$ & CP$_F$ & CP$_S$ \\[4pt]

				50
				& $\beta_1$
				& 0.007\he(0.7) & 0.188 & 0.187 & 0.174 & 0.953 & 0.925
				& -0.001\he(-0.1) & 0.233 & 0.234 & 0.215 & 0.954 & 0.926
				& -0.003\he(-0.3) & 0.322 & 0.320 & 0.289 & 0.952 & 0.915 \\

				& $\beta_2$
				& -0.003\he(1.4) & 0.170 & 0.170 & 0.157 & 0.951 & 0.919
				& -0.002\he(1.1) & 0.221 & 0.218 & 0.200 & 0.950 & 0.916
				& 0.001\he(-0.5) & 0.307 & 0.303 & 0.274 & 0.952 & 0.914 \\

				& $A(\tau\!/\!4)$
				& -0.024\he(-1.3) & 0.333 & 0.331 & 0.319 & 0.932 & 0.923
				& -0.049\he(-1.0) & 1.131 & 1.116 & 1.075 & 0.920 & 0.909
				& -0.096\he(-1.6) & 2.164 & 2.094 & 2.004 & 0.887 & 0.877 \\

				& $A(\tau\!/\!2)$
				& -0.018\he(-0.7) & 0.444 & 0.442 & 0.426 & 0.934 & 0.920
				& -0.048\he(-0.9) & 1.262 & 1.242 & 1.196 & 0.919 & 0.909
				& -0.096\he(-1.5) & 2.367 & 2.285 & 2.187 & 0.887 & 0.874 \\

				& $A(\tau)$
				& -0.009\he(-0.3) & 0.504 & 0.500 & 0.483 & 0.936 & 0.922
				& -0.047\he(-0.9) & 1.276 & 1.257 & 1.210 & 0.920 & 0.909
				& -0.096\he(-1.5) & 2.385 & 2.302 & 2.203 & 0.887 & 0.874 \\

				\addlinespace

				100
				& $\beta_1$
				& 0.003\he(0.3) & 0.130 & 0.129 & 0.125 & 0.949 & 0.936
				& 0.000\he(0.0) & 0.161 & 0.162 & 0.155 & 0.952 & 0.936
				& -0.001\he(-0.1) & 0.219 & 0.221 & 0.209 & 0.955 & 0.935 \\

				& $\beta_2$
				& 0.000\he(0.0) & 0.119 & 0.117 & 0.112 & 0.949 & 0.932
				& -0.002\he(0.8) & 0.150 & 0.150 & 0.144 & 0.951 & 0.937
				& -0.001\he(0.7) & 0.207 & 0.208 & 0.197 & 0.953 & 0.932 \\

				& $A(\tau\!/\!4)$
				& -0.007\he(-0.4) & 0.233 & 0.234 & 0.230 & 0.946 & 0.941
				& -0.005\he(-0.1) & 0.792 & 0.790 & 0.776 & 0.938 & 0.932
				& -0.021\he(-0.4) & 1.507 & 1.488 & 1.456 & 0.920 & 0.914 \\

				& $A(\tau\!/\!2)$
				& 0.003\he(0.1) & 0.313 & 0.311 & 0.307 & 0.947 & 0.941
				& 0.000\he(0.0) & 0.880 & 0.879 & 0.864 & 0.941 & 0.934
				& -0.022\he(-0.3) & 1.643 & 1.621 & 1.587 & 0.921 & 0.914 \\

				& $A(\tau)$
				& 0.014\he(0.5) & 0.354 & 0.353 & 0.347 & 0.946 & 0.941
				& 0.002\he(0.0) & 0.890 & 0.889 & 0.875 & 0.941 & 0.933
				& -0.023\he(-0.4) & 1.656 & 1.633 & 1.599 & 0.920 & 0.914 \\

				\addlinespace

				200
				& $\beta_1$
				& 0.003\he(0.3) & 0.091 & 0.091 & 0.089 & 0.953 & 0.942
				& -0.002\he(-0.2) & 0.113 & 0.113 & 0.111 & 0.954 & 0.946
				& 0.000\he(0.0) & 0.154 & 0.154 & 0.149 & 0.952 & 0.942 \\

				& $\beta_2$
				& 0.000\he(-0.2) & 0.082 & 0.082 & 0.080 & 0.951 & 0.945
				& -0.001\he(0.6) & 0.106 & 0.105 & 0.102 & 0.949 & 0.938
				& 0.002\he(-0.8) & 0.146 & 0.145 & 0.141 & 0.950 & 0.937 \\

				& $A(\tau\!/\!4)$
				& -0.002\he(-0.1) & 0.165 & 0.165 & 0.164 & 0.948 & 0.945
				& -0.008\he(-0.2) & 0.559 & 0.555 & 0.551 & 0.944 & 0.941
				& -0.021\he(-0.4) & 1.058 & 1.047 & 1.036 & 0.932 & 0.928 \\

				& $A(\tau\!/\!2)$
				& 0.010\he(0.4) & 0.220 & 0.220 & 0.219 & 0.947 & 0.944
				& -0.003\he(-0.1) & 0.623 & 0.618 & 0.613 & 0.945 & 0.940
				& -0.018\he(-0.3) & 1.157 & 1.143 & 1.130 & 0.932 & 0.928 \\

				& $A(\tau)$
				& 0.022\he(0.8) & 0.249 & 0.249 & 0.248 & 0.949 & 0.947
				& -0.002\he(0.0) & 0.631 & 0.626 & 0.620 & 0.943 & 0.941
				& -0.018\he(-0.3) & 1.164 & 1.151 & 1.139 & 0.932 & 0.928 \\

				\addlinespace

				400
				& $\beta_1$
				& 0.002\he(0.2) & 0.064 & 0.064 & 0.063 & 0.950 & 0.946
				& 0.001\he(0.1) & 0.081 & 0.080 & 0.079 & 0.948 & 0.944
				& 0.001\he(0.1) & 0.108 & 0.108 & 0.107 & 0.951 & 0.945 \\

				& $\beta_2$
				& -0.001\he(0.4) & 0.058 & 0.057 & 0.057 & 0.951 & 0.947
				& 0.001\he(-0.5) & 0.073 & 0.074 & 0.073 & 0.952 & 0.946
				& -0.001\he(0.6) & 0.102 & 0.102 & 0.101 & 0.950 & 0.946 \\

				& $A(\tau\!/\!4)$
				& 0.000\he(0.0) & 0.116 & 0.117 & 0.116 & 0.948 & 0.948
				& 0.010\he(0.2) & 0.392 & 0.393 & 0.392 & 0.950 & 0.948
				& 0.001\he(0.0) & 0.740 & 0.742 & 0.739 & 0.945 & 0.944 \\

				& $A(\tau\!/\!2)$
				& 0.011\he(0.4) & 0.155 & 0.155 & 0.155 & 0.950 & 0.949
				& 0.016\he(0.3) & 0.434 & 0.438 & 0.436 & 0.950 & 0.948
				& 0.005\he(0.1) & 0.806 & 0.809 & 0.806 & 0.946 & 0.944 \\

				& $A(\tau)$
				& 0.024\he(0.8) & 0.176 & 0.176 & 0.176 & 0.949 & 0.947
				& 0.016\he(0.3) & 0.440 & 0.443 & 0.441 & 0.950 & 0.949
				& 0.005\he(0.1) & 0.812 & 0.815 & 0.812 & 0.945 & 0.943 \\

			\end{tabular}
		}

		\vspace{2.5ex}

		\parbox{\linewidth}{\footnotesize Bias (Bias in \%), empirical standard deviation (SD), average estimated standard deviations based on inverse Fisher Information (SE$_F$) and Sandwich Estimator (SE$_S$), coverage probabilities (CP$_F$, CP$_S$). We investigated the following parameters and settings, with regression parameters $\beta = (1, -0.2)$ in all scenarios:\\[2ex]
			a) $r=0.5$, $\gamma_1=2.9$, $\gamma_2=0.8$, $\gamma_4=0.033$, $\tau=5$, $(A(\tau\!/4),\,A(\tau\!/2),\,A(\tau))=(1.833, 2.508, 2.847)$, avg. 2.0 recurrences, 13\% terminal events, 14\% censorings \\ \he b) $r=1$, $\gamma_1=5.3$, $\gamma_2=1.8$, $\gamma_4=0.025$, $\tau=5$, $(A(\tau\!/4),\,A(\tau\!/2),\,A(\tau))=(4.741, 5.241, 5.299)$, avg. 1.9 recurrences, 12\% terminal events, 15\% censorings \\ \he c) $r=2$, $\gamma_1=6.5$, $\gamma_2=2$, $\gamma_4=0.022$, $\tau=5$, $(A(\tau\!/4),\,A(\tau\!/2),\,A(\tau))=(5.967, 6.457, 6.500)$, avg. 1.3 recurrences, 12\% terminal events, 14\% censorings}
	\end{table}
\end{landscape}

Data were simulated sequentially with Gompertz-type cumulative baselines
$\Lambda_0^{\gamma_k,\gamma_\ell}(t)=\gamma_k\{1-\exp(-\gamma_\ell t)\}$. The first recurrent event was generated from $F_1(t|Z)=1-\exp\bigl[-\mathcal G\{e^{\beta_1^TZ}\,\Lambda_0^{\gamma_1,\gamma_2}(t)\}\bigr]$,
with $\varepsilon_{i1}\in\{1,2\}$ drawn from a Bernoulli distribution with parameter
$F_1(\infty|Z)$, and the terminal event analogously from $F_2(t|Z)$ with parameters $\beta_2$,
$\gamma_3$, $\gamma_4$; explicit expressions are given in Section \ref{sec:suppsim} of the Supplementary Material. Subsequent recurrences
$(t_{ij},\varepsilon_{ij})$ were generated from the conditional
distribution of $(T,\varepsilon)$ given $T>t_{i,j-1}$, i.e.\
$P(T\le t,\varepsilon=1\mid T>t_{i,j-1},Z)=P(t_{i,j-1}<T\le t,\varepsilon=1\mid Z)/P(T>t_{i,j-1}\mid Z)$.  

All simulation study results are based on $10\,000$ replicates (Tables~\ref{tab:sim_prop_rho} and~\ref{tab:sim_log_r_full}). The proposed estimator demonstrated strong finite-sample performance with regard to bias, standard errors, and coverage probabilities, and performed reasonably well even for sample size $n=50$, which is relevant for many clinical trials. SE$_F$ and SE$_S$ showed similar performance, although at $n = 50$ SE$_F$ was slightly closer to the empirical standard deviation. As the sandwich estimator remains valid under residual within-subject dependence, we recommend it for clinical study data.

We further investigated selection of the transformation by the Takeuchi and
Akaike information criteria, with data generated from Box--Cox models with $\rho\in\{0.2,0.5,1,2\}$ (Tables~\ref{tab:tic} and~\ref{tab:tic_cov}
in the Supplementary Material). The Takeuchi information criterion selected
transformations close to the true one, increasingly so with larger sample sizes,
and reliably detected departures from the Ghosh--Lin model, which was rejected in
at least 97\% of replicates for $n=400$ when misspecified. The Akaike
information criterion behaved similarly, but tended to overestimate the
transformation parameter in small samples. After selection, coverage of the
marginal mean was between 0.90 and 0.96 for $n=400$, whereas that of the
misspecified Ghosh--Lin model decreased with the sample size, to as low as 0.41
when the true transformation parameter was $\rho=2$.

\section{Application to the STATCOPE Interim Analysis Data}
\label{sec:statcope}
STATCOPE \citep{Criner.etal:2014} was a randomized, placebo-controlled trial evaluating simvastatin, in addition to usual care, for reducing exacerbations in patients with chronic obstructive pulmonary disease. 
The trial enrolled subjects in Global Initiative for Chronic Obstructive Lung Disease severity stages 2--4 (moderate, severe, and very severe disease). Death as a terminal event occurred in 6.7\% of the full cohort and 12.9\% of the stage 4 subgroup. The planned sample size of 1,200 subjects was based on a hypothesized 15\% reduction in exacerbations. To achieve the target event count, the trial design included 1,253 mild, self-reported exacerbations alongside 697 medically attended events involving emergency visits, hospitalizations, or intensive care admissions (Supplementary Material, Table~\ref{tab:statcope_exacerbations}). The trial was stopped early at 72.5\% enrollment based on a preplanned interim futility analysis. Using intention-to-treat negative binomial regression, the analysis found no treatment effect overall or in any prespecified subgroup. 

\begin{table}[ht]
\centering
\footnotesize
\setlength{\tabcolsep}{2.5pt}
\caption{\textsc{statcope} data, severe exacerbations: parameter estimates and standard errors for different regression models}
\label{tab:statcope_models}
\resizebox{\textwidth}{!}{%
\begin{tabular}{llrrrrrrr}
 & & \multicolumn{2}{c}{\textbf{Pseudo-likelihood: logarithmic}} & \multicolumn{2}{c}{\textbf{Pseudo-likelihood: Box--Cox}} & \textbf{Ghosh--Lin} & \textbf{Andersen--Gill} & \textbf{Neg.\ binomial}\\
\multirow{9}{*}{\rotatebox[origin=c]{90}{Stages 2--4}} &  & $r = 1$ & $r = 0.64$ & $\rho = 0.18$ & $\rho = 1$ &  &  & \\
 & $\Delta$TIC & 1.84 & 0.00 & 0.45 & 9.05 &  &  & \\
 & treat & $-$0.058 (0.187) & $-$0.085 (0.173) & $-$0.086 (0.178) & {\bf$-$0.125 (0.143)} & {\bf$-$0.126 (0.143)} & {\bf$-$0.110 (0.145)} & {\bf$-$0.011 (0.141)}\\
 & age70+ & $-$0.714 (0.256) & $-$0.637 (0.237) & $-$0.645 (0.244) & $-$0.434 (0.196) & $-$0.434 (0.197) & $-$0.458 (0.199) & $-$0.543 (0.200)\\
 & stage4 & 0.730 (0.189) & 0.650 (0.177) & 0.669 (0.182) & 0.471 (0.156) & 0.471 (0.156) & 0.511 (0.157) & 0.612 (0.144)\\
 & smoker & $-$0.338 (0.219) & $-$0.333 (0.205) & $-$0.339 (0.210) & $-$0.289 (0.173) & $-$0.289 (0.173) & $-$0.286 (0.174) & $-$0.297 (0.169)\\
 & pyear & 0.034 (0.011) & 0.031 (0.010) & 0.031 (0.011) & 0.023 (0.008) & 0.023 (0.008) & 0.025 (0.009) & 0.030 (0.008)\\
 & hosp & 1.684 (0.199) & 1.534 (0.182) & 1.578 (0.187) & 1.203 (0.151) & 1.204 (0.151) & 1.262 (0.151) & 1.336 (0.145)\\
 & oxygen & 0.433 (0.205) & 0.384 (0.192) & 0.397 (0.196) & 0.280 (0.167) & 0.281 (0.167) & 0.327 (0.168) & 0.424 (0.158)\\
\multirow{8}{*}{\rotatebox[origin=c]{90}{Stage 4}} &  & $r = 1$ & $r = 0.98$ & $\rho = 0.06$ & $\rho = 1$ &  &  & \\
 & $\Delta$TIC & 0.07 & 0.07 & 0.00 & 4.23 &  &  & \\
 & treat & $-$0.136 (0.280) & $-$0.137 (0.279) & $-$0.143 (0.274) & {\bf$-$0.156 (0.196)} & {\bf$-$0.156 (0.197)} & {\bf$-$0.101 (0.201)} & {\bf$-$0.044 (0.193)}\\
 & age70+ & $-$0.500 (0.390) & $-$0.496 (0.388) & $-$0.475 (0.381) & $-$0.253 (0.271) & $-$0.253 (0.271) & $-$0.268 (0.275) & $-$0.388 (0.271)\\
 & smoker & 0.240 (0.315) & 0.238 (0.313) & 0.227 (0.306) & 0.119 (0.211) & 0.120 (0.212) & 0.100 (0.221) & 0.111 (0.208)\\
 & pyear & 0.014 (0.014) & 0.014 (0.014) & 0.014 (0.014) & 0.009 (0.009) & 0.009 (0.009) & 0.012 (0.009) & 0.017 (0.009)\\
 & hosp & 1.367 (0.279) & 1.360 (0.278) & 1.335 (0.274) & 0.917 (0.201) & 0.917 (0.202) & 0.989 (0.205) & 1.068 (0.191)\\
 & oxygen & 0.704 (0.293) & 0.698 (0.292) & 0.672 (0.286) & 0.373 (0.212) & 0.374 (0.212) & 0.424 (0.217) & 0.620 (0.219)\\
\end{tabular}}
\par\vspace{1mm}\parbox{\textwidth}{\footnotesize
Columns 3--6: pseudo-likelihood estimator under logarithmic and Box--Cox transformations, including the proportional odds model ($r = 1$), models selected by the Takeuchi information criterion ($r = 0.64$, $\rho = 0.18$ for stages 2--4; $r = 0.98$, $\rho = 0.06$ for stage 4), and the identity link ($\rho = 1$); $\Delta$TIC denotes the difference in the Takeuchi information criterion relative to the best-fitting model in each subset. Columns 7--9: Ghosh--Lin weighted estimating equation, Andersen--Gill model, and negative binomial regression with log person-years as offset. Parameter estimates with sandwich standard errors in parentheses. Covariates: treat (simvastatin), age70+ (age $\geq 70$), stage4 (severity stage 4), smoker (current smoker), pyear (pack-years), hosp (hospitalization for COPD in the previous year), oxygen (supplemental oxygen at baseline).}
\end{table}

We restricted our analysis to exacerbations of severity 3--4, which required hospital admission, retaining 436 of 1,950 exacerbations (22.4\%) in the full cohort (stages 2--4) and 227 of 821 (27.6\%) in the stage 4 subset. Our reanalysis is therefore underpowered for statistical significance. Covariates were selected from baseline characteristics by backward elimination under the identity transformation, removing covariates with $p > 0.10$ while retaining treatment; the stage 4 subset uses the same covariates. We fitted our pseudo-likelihood model under logarithmic and Box--Cox transformations to both cohorts to estimate covariate effects on the marginal rate of severe exacerbations (Table~\ref{tab:statcope_models}). Under the identity transformation, our estimates closely match those from the Ghosh--Lin model, fitted with the R package \texttt{mets} \citep{Scheike:2025}, with only minor numerical deviations, confirming that their weighted estimating equation is a special case of our approach (Section~\ref{sec:ghosh-lin}). We also compared our estimates with those from the Andersen--Gill model \citep{Andersen.etal:1982} (equivalently the Zeng--Lin model with identity transformation, \citealp{ZengLin:2006}) and negative binomial regression.

\begin{figure}[b!]
	\centering
	\includegraphics[scale=0.8]{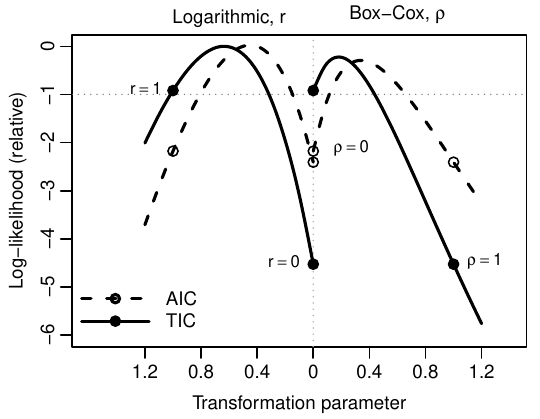}
	\vspace{-2ex}
	\caption{\footnotesize Model selection using the Takeuchi and Akaike information criteria (TIC, AIC) for GOLD stages 2--4. Profile log-pseudo-likelihood curves are shown for the logarithmic transformation models (left side) and Box--Cox transformation models (right side).}
	\label{fig:loglik}
\end{figure}

\begin{figure}[t!]
	\centering
	\includegraphics[scale=0.6]{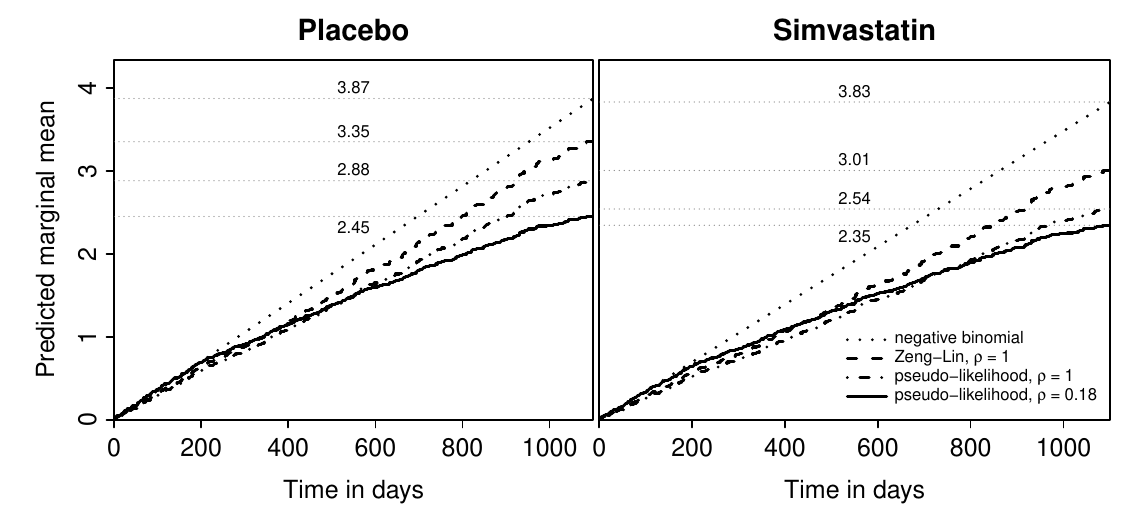}
	\vspace{-1ex}
	\caption{\footnotesize Individualized prediction of the marginal mean of severe exacerbations
		for a former smoker aged under 70 with stage 4 disease, 40 pack-years, prior
		hospitalization and supplemental oxygen, assigned to placebo (left panel) or
		simvastatin (right panel), under negative binomial regression, the Zeng--Lin
		model ($\rho=1$) and the pseudo-likelihood approach (Box--Cox, $\rho=1$ and $\rho=0.18$).}
  	\label{fig:predict_trans}
\end{figure}

Under the identity transformation, the pseudo-likelihood approach and the Ghosh--Lin model estimate an 11.8\% reduction in the marginal rate of severe exacerbations under simvastatin in the full cohort ($\exp(-0.125) = 0.88$, 95\% confidence interval 0.67 to 1.17) and a 14.4\% reduction among stage 4 patients (Table~\ref{tab:statcope_models}), of a similar magnitude to the 15\% reduction assumed in the trial design. The Andersen--Gill model, which describes the rate among surviving patients, gives reductions of 10.4\% and 9.6\%, whereas negative binomial regression, the method of the interim analysis, shows almost no effect (1.1\% and 4.3\%). In all models, the confidence intervals are wide because of the reduced sample size, and a larger trial, powered for severe exacerbations, would be needed to confirm a treatment effect.

Figure~\ref{fig:loglik} displays the profile log-pseudo-likelihood for different
transformation parameters across logarithmic transformation models (left side)
and Box--Cox transformation models (right side). Among the Box--Cox models,
$\rho=0.18$ is optimal by the Takeuchi information criterion, while $r=0.64$
yields the best fit among the logarithmic models. The Ghosh--Lin model
($\rho=1$) provides a clearly poorer fit ($\Delta$TIC $=9.05$, $\Delta$AIC $=4.81$;
Table~\ref{tab:statcope_models}). In the stage 4 subset, both families select
models close to the proportional odds model ($\rho=0.06$ and $r=0.98$; the
proportional odds model corresponds to $\rho=0$ and $r=1$), while the Ghosh--Lin
model again provides a poorer fit ($\Delta$TIC $=4.23$, $\Delta$AIC $=1.08$).

Figure~\ref{fig:predict_trans} shows the predicted marginal mean of severe
exacerbations for a former smoker aged under 70 with stage 4 disease,
40 pack-years, prior hospitalization and supplemental oxygen. Under the identity
transformation, $\exp(\beta)$ is the ratio of the marginal means for every
subject and time point, whereas under the other transformations the treatment
effect varies with the subject's risk and over time. The predicted treatment
benefit is model-dependent: at 1100 days, the Ghosh--Lin model predicts 2.88
exacerbations under placebo and 2.54 under simvastatin (11.8\% fewer), whereas
the Box--Cox model with $\rho=0.18$, selected by the Takeuchi information
criterion, predicts 2.45 and 2.35 (4.3\% fewer). The Zeng--Lin model and negative
binomial regression treat death as censoring and overestimate the marginal mean,
with 3.35 and 3.87 exacerbations under placebo; negative binomial regression,
which assumes a constant event rate, shows almost no difference between the
treatment arms (3.83 under simvastatin).

\section{Discussion}\label{sec:discuss}
We propose a general semiparametric regression framework for the marginal mean of recurrent events with a competing terminal event. Modeling the marginal rate through a weighted pseudo-likelihood function, we develop a direct approach that does not rely on the dependence structure between the recurrent and terminal events, and includes the Ghosh--Lin model as a special case. While external time-dependent covariates and left-truncated pseudo-subjects capture part of the within-subject dependence, residual unobserved dependence does not affect consistency and is accounted for by the sandwich variance estimator. Modeling this residual dependence through a frailty term is of interest for future work and may improve the efficiency of the estimators. Neural network components are another promising extension for large datasets such as
registries or electronic health records. For clinical trials, our approach
already accounts for event times, censoring, and competing terminal events, in
contrast to count-based methods, and performs reasonably well even for $n=50$. We therefore
expect that semiparametric models for the marginal rate will continue to play an important role in the analysis of clinical study data.

For the STATCOPE interim analysis data, we demonstrated that, under the identity
transformation, covariate effects and predictions from the Zeng--Lin model and
negative binomial regression differ from those of the pseudo-likelihood approach,
and would therefore be misleading if interpreted in terms of the marginal mean.
In Section~B.6 of the Supplementary Material, we present a second data example on
hospital readmissions, with 27\% deaths, where these differences are substantial.
The original endpoint of the study combined mild, self-reported exacerbations with
medically attended events involving emergency visits, hospitalizations, or
intensive care admissions, which may have diluted treatment effects on the
clinically more relevant components. Because 12.9\% of stage 4 patients died
during follow-up, an analysis of severe exacerbations should account for death as
a competing terminal event. Marginal rate models, such as the pseudo-likelihood
approach proposed here, combined with an analysis of mortality, appear to be
well suited for this purpose. Interpreting composite endpoints when treatment
effects act in opposing directions across components remains challenging
\citep{Song.etal:2008}. Weighting components by clinical importance, as proposed
by \citet{MaoLin:2016}, could be explored within our pseudo-likelihood framework
in future research.

\section{Acknowledgement}\label{sec:6}
Part of the research of the first author was conducted at the National Heart, Lung, and Blood
Institute, U.S. National Institutes of Health. Claude (Anthropic) was used for text editing,
grammar correction, code development and figure refinement; the authors reviewed and edited
the content and take full responsibility for the publication. The authors have no conflicts of
interest to declare.

\section*{Supplementary material}
The Supplementary Material contains details on the calculation of the estimates, a lemma for
weighted $Z$-estimators, details on the variance estimation, details on the data generation and
simulation results for the model selection, and additional results for the STATCOPE trial.
The proposed estimator is implemented in the R package \texttt{wnpmle}, available on CRAN and
GitHub; simulation code is available from the corresponding author. The STATCOPE trial data
(ClinicalTrials.gov: NCT01061671) were obtained from the trial investigators and are not
publicly available; access requests should be directed to them.

\appendix
\section{Appendix}\label{Appendix}
\subsection{Model conditions}\label{sec:Acondition}
	\begin{list}{}{\setlength{\leftmargin}{2.3em}\setlength{\labelwidth}{1.8em}}	
	\item[M1)] The cumulative baseline $\Lambda_0(t)$ is a strictly
	increasing and continuously differentiable function and $\beta_{_0}$
	lies in the interior of a compact set $\mathcal C$.
	\item[M2)] The vector of covariates $Z(t)$ is $P$-almost surely of
	bounded variation on the observed interval $[0,\tau]$. In combination with model condition M1) this implies that there is a constant $M$ such that $\sup_{\beta\in\mathcal C,t\in[0,\tau]}|\beta^TZ(t)|\leq M$.
	\item[M3)] The endpoint of the study $\tau$ is chosen in a way that $P$-almost surely there exists a constant $\delta>0$ such that $P(C\geq \tau|Z)>\delta$ and $P(X\geq\tau|Z)>\delta$. 
	\item[M4)] $\mathcal G$ is a thrice continuously differentiable and strictly increasing function with $\mathcal G(0)=0,\he\mathcal G'(0)>0$ and $\mathcal G(\infty)=\infty$. One of the following additional conditions is required:
	\begin{itemize}
		\item[$a)$] $\mathcal G''(x)\leq 0$\, for \,$x>0$ or
		\item[$b)$] $\mathcal G''(x)\geq 0$\, for $x>0$. In addition to that for any $a\in(0,\infty)$ and for any sequence $(x_n)\subset\mathbb{R}$ with $x_n\to\infty$ as $n\to\infty$, 
		\begin{equation*}
		(\ast)\hee\lim_{n\to\infty}\mathbb P(X\leq\tau,\Delta\eps=1)\left\{\frac{\log(ax_n)}{\mathcal G(a^{-1}x_n)}+\frac{\log\,\mathcal G'(ax_n)}{\mathcal G(a^{-1}x_n)}\right\}<P(X\geq\tau).\hspace{3cm}
		\end{equation*}
	\end{itemize}
	\item[M5)] {\em Identifiability condition}. If a vector $h_1\in\mathbb{R}^d$ and a deterministic function $h_2(t),\,t\in[0,\tau]$ exist such that $h_1^TZ(t)+h_2(t)=0$ $P^{\theta_0}$-almost surely, then $h_1=0$ and
	$h_2(t)=0$, for $\mu$ almost all $t\in [0,\tau]$, with $\mu$ denoting the  Lebesgue measure. This condition rules out  perfect multicollinearity of $Z_1(t),\ldots,Z_k(t)$ for $\mu-$ almost all $t\in[0,\tau]$.
	\item[M6)] Let $h_1$ denote a vector in $\mathbb{R}^d$ and let $h_2(t)\in\mathcal D$.
	For a subset $\mathcal S\subset[0,\tau]$ of nonzero Lebesgue measure it is required that $\forall t\in\mathcal S$
	\begin{equation*}
	P^{\theta_0}\left[h_1^TZ(t)+h_2(t)\neq-\,\frac{\mathcal G''\left\{\int_0^t e^{\beta_0^TZ(u)}d\Lambda_0(u)\right\}}{\mathcal G'\left\{\int_0^t e^{\beta_0^TZ(u)}d\Lambda_0(u)\right\}}\int_0^te^{\beta_0^TZ(u)} \left\{h_1^TZ(u)+h_2(u)\right\}d\Lambda_0(u)\right]>0.
	\end{equation*}
\end{list}
\subsection{Existence and Consistency of the Pseudo-Likelihood Estimator}
\label{sec:Aconsist}
Extending arguments of \citet{Murphy.etal:1994} and of \citet*{ZengLin:2006} we establish existence and consistency of the  proposed pseudo-likelihood estimator for the marginal rate of recurrent events with a competing terminal event under general semiparametric transformation models. 

For existence, the key argument is to show that if $\widehat{\Lambda}_n(\tau)$ was unbounded, $n^{-1}[\ell(\widehat{\Lambda}_n,\widehat{\beta}_n)-\ell(\widetilde{\Lambda}_n,\beta_0)]\to-\infty$ for some function $\widetilde{\Lambda}_n(t)$, which contradicts the definition of $\widehat{\Lambda}_n$ as a pseudo-likelihood estimator. Following the approach by \citet{Bellach.etal:2019} we define $\widetilde{\Lambda}_n(t):=n^{-1}\sum_{i=1}^n N_i(t)$ to establish the contradiction under weak assumptions. By Jensen's inequality it is obtained that
$n^{-1}\bigl[\ell(\widehat{\Lambda}_n,\widehat{\beta}_n)-\ell(\widetilde{\Lambda}_n,\beta_0)\bigr]$
is bounded from above under model condition M4a) by
\begin{equation*}
\left[\mathbb{P}_n\mathds{1}\{X\leq\tau,\Delta\eps=1\}\,\frac{\log\,\mathcal G\bigl(e^{-M} 
		\widehat{\Lambda}_n(\tau)\bigr)}{\mathcal G\bigl(e^{-M} 
		\widehat{\Lambda}_n(\tau)\bigr)}-\mathbb{P}_n\mathds{1}\{X\geq\tau\}\right]
\mathcal G\bigl(e^{-M}\widehat{\Lambda}_n(\tau)\bigr)+O_p(1),
\end{equation*}
and under model condition M4b) by 
\begin{eqnarray*}
&&\left[\mathbb{P}_n\mathds{1}\{X\leq\tau,\Delta\eps=1\}\left[\frac{\log\,\mathcal G'\bigl(e^M\widehat{\Lambda}_n(\tau)\bigr)}{\mathcal G\bigl(e^{-M}\widehat{\Lambda}_n(\tau)\bigr)}
+\frac{\log\bigl(e^M\widehat{\Lambda}_n(\tau)\bigr)}{\mathcal G\bigl(e^{-M}\widehat{\Lambda}_n(\tau)\bigr)}\right]-\mathbb{P}_n\mathds{1}\{X\geq\tau\}\right]\\[1ex]
&&\hspace{8cm}\times\mathcal G\bigl(e^{-M}\widehat{\Lambda}_n(\tau)\bigr)+O_p(1).
\end{eqnarray*}	
The contradiction is readily established in both cases. As a consequence, $\widehat{\Lambda}_n(\tau)$ is almost surely bounded and the estimated jump sizes must be finite.
\citet{Bellach.etal:2019} show that M4a) is satisfied for logarithmic transformation models and for  Box--Cox transformation models with $\rho\in[0,1]$, while M4b) holds for Box--Cox transformation models with $\rho>1$.

With the Helly--Bray Lemma it is obtained that for every sequence $\widehat{\Lambda}_n$ there is a subsequence $\widehat{\Lambda}_{n_k}$ with $\widehat{\Lambda}_{n_k}\to \Lambda_*$ for some function $\Lambda_*$. The parameter space is denoted by\, $\Theta=\{\theta=(\beta,\Lambda)$ with $\beta\in\mathbb R^d$ and $\Lambda$ being a monotone increasing element of $\mathcal D[0,\tau]\}$. Sequential compactness for $\Theta$ is then obtained under model condition M1). As $\Lambda_{_0}$ is a continuous and increasing function, we obtain convergence uniformly almost surely for the cumulative baseline hazard $\widehat{\Lambda}_n$ and almost sure convergence for $\widehat{\beta}_n$.  
\\[1ex]
{\em Recurrent and Cure Events}. Differentiating the log-likelihood with respect to jump sizes we obtain
\begin{equation*}
\widehat{\Lambda}_n(t)=\int_0^t \frac{\frac{1}{n}\sum_{i=1}^n 
		\mathds{1}(C_i\geq s)dN_i(s)}{|\Phi_n^c(s,\widehat{\Lambda}_n,\widehat{\beta}_n)|}
\end{equation*}
with 
\begin{equation*}
	\Phi_n^c(s,\Lambda,\beta) = \frac{1}{n}\sum_{i=1}^n \mathds{1}(C_i\geq s)
	Y_i(s)e^{\beta^TZ_i(s)}\mathcal{G}'\left(\int_0^s e^{\beta^TZ_i(u)}d\Lambda(u)\right).
\end{equation*}
Substituting estimated parameters by the true model parameters and
applying the decomposition $dN_i(t) = \lambda(t)Y_i(t)\,dt + dM_i(t)$
with $M_i$ of mean zero, by the Glivenko--Cantelli theorem
$\widetilde\Lambda^c_n(t) \to \Lambda_0(t)$ uniformly almost surely.
Further, $\Phi_n^c(s,\widehat{\Lambda}_n,\widehat{\beta}_n)$ 
converges uniformly to a continuously differentiable function 
$\Phi_*^c(s,\Lambda_*,\beta_*)$ bounded away from zero, and
\begin{equation*}
	\widehat{\Lambda}_n(t)\to\Lambda_*(t)=\int_0^t
	\frac{E\bigl[\eta(s,\beta_0,\Lambda_0)\bigr]}{\Phi_*^c(s,\Lambda_*,\beta_*)}d\Lambda_0(s),
\end{equation*}
with $\eta(s,\beta_0,\Lambda_0)=\mathds{1}(C\geq s)Y(s)e^{\beta_0^TZ(s)}
\mathcal{G}'\left(\int_0^s e^{\beta_0^TZ(u)}d\Lambda_0(u)\right)$.

With a Kullback--Leibler argument we show that every subsequence converges to
the true parameter $\theta_0 = (\beta_0, \Lambda_0)$. This is first
established for the model with recurrent and cure events, where censoring
is observable after the cure event. The Kullback--Leibler information is
defined as
\[
\mathcal{K}_c(\theta_0, \theta_*) = \int E_{\theta_0}\left[\log\left\{
\prod_{j}\left(\frac{\lambda_{\theta_0}(X_j \mid Z)}{\lambda_{\theta_*}(X_j \mid Z)}\right)^{\Delta_j 1(\varepsilon_j = 1)}
\frac{S_{\theta_0}(C \wedge \tau \mid Z)}{S_{\theta_*}(C \wedge \tau \mid Z)}
\right\}\right] dG_c(c),
\]
for $\theta_0 \equiv (\beta_0, \Lambda_0)$ and $\theta_* \equiv (\beta_*,
\Lambda_*)$, where $X_j$ denote the recurrence times of a subject and
$S_\theta(t \mid Z) = \exp\{-\Lambda_\theta(t \mid Z)\}$. The term inside the
logarithm is the ratio of the likelihood contributions in
Section~\ref{sec:curewcens}. Since the pseudo-log-likelihood is linear in
$dN_i$, its expectation depends on the distribution of the recurrent events
only through the marginal rate; $\mathcal{K}_c$ is therefore the limit
of the expected pseudo-log-likelihood ratio, and the argument does not
require the working assumption, as in \citet{Linetal:2000} and
\citet{WellnerZhang:2007}. With arguments as in \citet{Bellach.etal:2019} it
is then established that $\mathcal{K}_c(\theta_0, \theta_*) \le 0$ and
$\mathcal{K}_c(\theta_0, \theta_*) \ge 0$, and identifiability,
$\mathcal{K}_c(\theta_0, \theta_*) = 0 \Rightarrow \theta_0 = \theta_*$.
\\[1ex]
{\em Recurrent and Absorbing Events with Independent Right Censoring}.
From maximizing the likelihood function with regard to jump sizes we obtain
\[
\widehat{\Lambda}_n(t)=\int_0^t\frac{\frac{1}{n}\sum_{i=1}^n 
		\mathds{1}(C_i\geq s)dN_i(s)}{|\Phi_n^\omega(s,\widehat{\Lambda}_n,\widehat{\beta}_n)|},
\]
where\,
	$\Phi_n^\omega(s,\Lambda,\beta) = \frac{1}{n}\sum_{i=1}^n w_i(s)
	Y_i(s)e^{\beta^TZ_i(s)}\mathcal{G}'\bigl(\int_0^s e^{\beta^TZ_i(u)}d\Lambda(u)\bigr)$.
\\[1ex]
Since $\Phi_n^\omega(s,\widehat{\Lambda}_n,\widehat{\beta}_n)=
\Phi_n^c(s,\widehat{\Lambda}_n,\widehat{\beta}_n)+o_p(1)$, we obtain 
$\widehat{\Lambda}_n(t)\to\Lambda_0(t)$ uniformly almost surely. By the 
Glivenko--Cantelli theorem $\Phi_n^\omega(s,\widehat{\Lambda}_n,\widehat{\beta}_n)$ 
converges uniformly to a continuously differentiable function 
$\Phi_*^\omega(s,\Lambda_*,\beta_*)$ bounded away from zero, and
\[
	\widehat{\Lambda}_n(t)\to\Lambda_*(t)=\int_0^t
	\frac{E\bigl[\eta(s,\beta_0,\Lambda_0)\bigr]}{\Phi_*^\omega(s,\Lambda_*,\beta_*)}
	d\Lambda_0(s).
\]
The Kullback--Leibler information is
\[
\begin{aligned}
	\mathcal{K}_{\omega^*}(\theta_0, \theta_*) = \int E_{\theta_0}\Biggl[\log\Biggl\{
	&\prod_{j}\left(\frac{\lambda_{\theta_0}(X_j \mid Z)}{\lambda_{\theta_*}(X_j \mid Z)}\right)^{\Delta_j 1(\varepsilon_j = 1)}
	\left(\frac{S_{\theta_0}(C \wedge \tau \mid Z)}{S_{\theta_*}(C \wedge \tau \mid Z)}\right)^{1(C < D)} \\
	&\times \left(\frac{S_{\theta_0}(D \wedge \tau \mid Z)\,\exp\{-\int_D^\tau \omega^*(t)\,d\Lambda_{\theta_0}(t \mid Z)\}}
	{S_{\theta_*}(D \wedge \tau \mid Z)\,\exp\{-\int_D^\tau \omega^*(t)\,d\Lambda_{\theta_*}(t \mid Z)\}}\right)^{1(D \le C)}
	\Biggr\}\Biggr] dG_c(c).
\end{aligned}
\]
We show that it is asymptotically equivalent to $\mathcal{K}_c(\theta_0,\theta_*)$ 
for the model with recurrent and cure events. Equivalence is obtained by
\begin{align*}
	&\left|\mathcal{K}_{\omega^*}(\theta_0,\theta_*)-\mathcal{K}_c(\theta_0,\theta_*)\right|\\[1ex]
	&\quad=\!\left|\int\! E_{\theta_0}\!\left[\mathds{1}(D\leq c)\log\!\left\{\!
	\frac{\exp(-\int_D^\tau w^*(t)d\Lambda_{\theta_0}(t))}
	{\exp(-\int_D^\tau w^*(t)d\Lambda_{\theta_*}(t))}
	\Big\slash
	\frac{\exp(-\int_D^\tau \mathds{1}(c\geq t)d\Lambda_{\theta_0}(t))}
	{\exp(-\int_D^\tau\mathds{1}(c\geq t)d\Lambda_{\theta_*}(t))}
	\!\right\}\right]dG_c(c)\right|\\[1ex]
	&\quad\leq\int_D^\tau\!\int E_{\theta_0}\left|w^*(t)-\widetilde{w}(t)\right|\,
	|d(\Lambda_{\theta_*}-\Lambda_{\theta_0})(t)|\,dG_c(c)\\[1ex]
	&\qquad+\int_D^\tau\!\int E_{\theta_0}\left|\widetilde{w}(t)-\mathds{1}(c\geq t)\right|\,
	|d(\Lambda_{\theta_*}-\Lambda_{\theta_0})(t)|\,dG_c(c)=o_p(1).
\end{align*}
Consistency then follows from the asymptotic equivalence of 
$\mathcal{K}_c(\theta_0,\theta_*)$ and $\mathcal{K}_{\omega^*}(\theta_0,\theta_*)$.

\subsection{Weak Convergence to a Gaussian Process}
\label{sec:Appendweak} 
To establish weak convergence we verify the conditions of the Lemma weighted $\mathcal{Z}$-estimators (\citet{Bellach.etal:2019}) in the Supplementary Material. This lemma extends Theorem 3.3.1 of \citet*{VanderVaart.etal:2023} to settings with random weight functions.

Let $\mathcal H=\{h=(h_1,h_2):\,h_1\in\mathbb{R}^d,\,h_2\in\mathcal D[0,\tau]\}$ with norm $\|h\|_{\mathcal H}=\|h_1\|+\|h_2\|_v$, and $\mathcal H_p=\{h\in\mathcal H:\,\|h\|_{\mathcal H}\leq p\}$ for $p<\infty$. The empirical score operator is $\Psi_n(\theta,h,w)=\mathbb{P}_n\psi(\theta,h,w)$, with components $\mathbb{P}_n\psi_1$ and $\mathbb{P}_n\psi_2$ corresponding to derivatives with respect to $\beta$ and $\Lambda$ respectively. The limiting version $\Psi$ replaces $\mathbb{P}_n$ by $\mathcal P$.
The first condition of the Lemma follows immediately from Donsker preservation theorems, see \citet*{VanderVaart.etal:2023} and \citet{Kosorok:2008}. The second condition is valid because pointwise convergence can be strengthened to $\mathcal L^2$ convergence by dominated convergence. By the Donsker theorem $\sqrt{n}(\Psi_n-\Psi)(\theta_{_0},\widetilde{w})$ converges in distribution to a tight random element $\mathcal Z_1$, and $\sqrt{n}\big\{(\Psi_n-\Psi)(\theta_{_0},\widehat{w}_n)-(\Psi_n-\Psi)(\theta_{_0},\widetilde{w})\big\}$ converges to a tight random element $\mathcal Z_2$. Joint convergence follows from asymptotic linearity of the two components and the fact that the composition of two Donsker classes is Donsker. By definition $\Psi_n(\widehat{\theta}_n,\widehat{w}_n)=0$, and $\Psi(\theta_0,\widetilde{w})=0$ follows from the non-negativity of the Kullback--Leibler information by interchanging expectation and differentiation, as in \citet*{Parner:1998}.

Fréchet differentiability follows from Gateaux differentiability with an additional continuity condition. Invertibility of $\dot{\Psi}_0^a$ for the model with recurrent and cure events can be derived with arguments as in \citet{Parner:1998}, \citet{ZengLin:2006} and \citet{Bellach.etal:2019}. For the model with recurrent and terminal events, the scores differ only in that $\mathds{1}(C_i\geq t)$ is replaced by $w^*(t)$. By the Glivenko--Cantelli theorem $\sup_t|w^*(t)-\mathds{1}(C_i\geq t)|=o_p(1)$, and therefore $\dot{\Psi}_0=\dot{\Psi}_0^a+o_p(1)$. Continuous invertibility of $\dot{\Psi}_0$ then follows from continuous invertibility of $\dot{\Psi}_0^a$. 

From Lemma for weighted $\mathcal Z-$estimators	we then obtain weak convergence of \\ $\sqrt{n}(\hat{\theta}_n-\theta_0)$ and the covariances for the limiting process \,$-\dot{\Psi}_0^{-1}(\mathcal Z_1+\mathcal Z_2)$, with $\mathcal Z_1(h)=\lim_{n\to\infty}\mathbb{P}_n\,\psi(\widehat{\beta}_n,\widehat{\Lambda}_n,h,\widetilde{w})$
and $\mathcal Z_2(h)=\lim_{n\to\infty}\mathbb{P}_n\big\{\psi(\widehat{\beta}_n,\widehat{\Lambda}_n,h,\widehat{w}_n)-\psi(\widehat{\beta}_n,\widehat{\Lambda}_n,h,\widetilde{w})\big\}$ are
\begin{center}
	$\mathcal P\left(\left[(\mathcal Z_1+\mathcal Z_2)\big\{\dot{\Psi}^{-1}_0(g)\big\}
	\right]\left[(\mathcal Z_1+\mathcal Z_2)\big\{\dot{\Psi}^{-1}_0(h)\big\}\right]\right)$
\end{center}
for $g,h\in\mathcal H$ (\citet{Kosorok:2008}). 


\clearpage
\section{Supplementary Material}
\subsection{Calculation of Estimates}
\label{sec:calc}
For Recurrent and Absorbing Events with Independent Right-Censoring the weighted log-likelihood function is approximated with 
\begin{eqnarray*}
	\ell_n&=&\sum_{i=1}^n\,\,\sum_{j=1}^{n_i}\left[\log \Lambda_n\{X_{ij}\}+\beta^TZ_i(X_{ij})+\log
	\mathcal G'\Big(\sum_{\stackrel{k,r: X_{kr}\leq X_{ij},}{\Delta_{kr}\eps_{kr}=1}}e^{\beta^T
		Z_i(X_{kr})} \Lambda_n\{X_{kr}\}\Big)\right]\\[0.2ex]
	&&-\he \sum_{i=1}^n\left[
	\mathcal G\Big(\sum_{k,r:\Delta_{kr}\eps_{kr}=1}\mathds{1}\bigl((D_i\wedge C_i\wedge\tau)\geq X_{kr}\bigr)e^{\beta^T
		Z_i(X_{kr})}\Lambda_n\{X_{kr}\}\Big)\,\right]\\[0.2ex]
	&&-\he \sum_{i:C_i>D_i}\left[\sum_{k,r:\Delta_{kr}\eps_{kr}=1}w_i^*(X_{kr})
	\mathds{1}(D_i<X_{kr})e^{\beta^TZ_i(X_{kr})}\Lambda_n\{X_{kr}\}\right.\\[-0.5cm]
	&&\hspace{6cm}\times\left. \mathcal G'\bigl(\sum_{\stackrel{m,s: X_{ms}\leq X_{kr},}{\Delta_{ms}\eps_{ms}=1}}e^{\beta^TZ_i(X_{ms})}\Lambda_n\{X_{ms}\}\bigr)\right]
\end{eqnarray*}
\normalsize
From maximizing $\ell_n$ we obtain estimates $\widehat{\beta}_n$ and 
$\widehat{\Lambda}_n\{X_{ij}\}$ for $i=1,\ldots,n$ and $j=1,\ldots,n_i$.

\subsection{Lemma for Weak Convergence of Weighted $\mathcal{Z}$--Estimators}
\label{sec:lemma3}

\begin{lemma}\label{lemma3} Let the parameter set $\Theta$ be a
	subset of a Banach space. Let $\widetilde{w}(t)$ be a
	bounded deterministic weight function and let $\widehat{w}_n(t)$ be a sequence of bounded random weight functions with values in $\mathbb
	R_+$. Let $\Psi_n$ and $\Psi$ be a linear random
	map and a linear deterministic map, respectively from $\Theta\times \mathbb R_+$ into a Banach space such that 
	\begin{center}
		$\displaystyle \hee \sqrt{n}(\Psi_n-\Psi)(\widehat{\theta}_n,\widehat{w}_n)-\sqrt{n}(\Psi_n-\Psi)(\theta_0,\widehat{w}_n)=o_p^*\bigl(1+\sqrt{n}\|\widehat{\theta}_n-\theta_0\|\bigr)$,
	\end{center}
	$\sqrt{n}(\Psi_n-\Psi)(\theta_0,\widetilde{w})$
	converges to a tight limit $\mathcal Z_1$ and 
	$\sqrt{n}\big((\Psi_n-\Psi)(\theta_0,\widehat{w}_n)-(\Psi_n-\Psi)(\theta_0,\widetilde{w})\big)$ converges to
	a tight limit $\mathcal Z_2$, and the sequences jointly converge to $(\mathcal Z_1,\mathcal Z_2)$.
	
	Let $(\theta,w)\to\Psi(\theta,w)$ be
	Fr\'echet-differentiable at $(\theta_0,\widetilde{w})$ with a continuously invertible
	derivative $\dot{\Psi}^{\widetilde{w}}_{\theta_0}$. If \,$\Psi(\theta_0,\widetilde{w})=0$ and
	$\widehat{\theta}_n$ satisfies $\Psi_n(\widehat{\theta}_n,\widehat{w}_n)=o_p^*(n^{-1/2})$, if  $\Psi_n(\widehat{\theta}_n,\widetilde{w})=\Psi_n(\widehat{\theta}_n,\widehat{w}_n)+o_p(1)$
	and if $\Psi_n(\widehat{\theta}_n,\widetilde{w})$ converges in outer probability to $\Psi(\theta_0,\widetilde{w})$, then
	\\[-3ex]
	\begin{equation*}\label{vw331}
		\sqrt{n}\,\dot{\Psi}_{\theta_0}^{\widetilde{w}}(\widehat{\theta}_n-\theta_0)=
		-\sqrt{n}\bigl((\Psi_n-\Psi)(\theta_0,\widetilde{w})+(\Psi_n-\Psi)(\theta_0,\widehat{w})-(\Psi_n-\Psi)(\theta_0,\widetilde{w})\bigr)+o_p^*(1)
	\end{equation*}
	and\, $\sqrt{n}(\widehat{\theta}_n-\theta_0)\leadsto-\bigl[\dot{\Psi}_{\theta_0}^{\widetilde{w}}\bigr]^{-1}(\mathcal Z_1+\mathcal Z_2)$, with
	$\leadsto$ denoting weak convergence.
\end{lemma}

\subsection{Variance Estimation}
\label{sec:varest}
We extend the theory developed in Bellach et al.\ (2019, 2020) for the subdistribution of a competing risk to settings with recurrent and terminal events. Arguments of Murphy (1994, 1995), Parner (1998), and Zeng \& Lin (2006) are utilized throughout. The middle part of the sandwich estimator is obtained from the iid decomposition of the score with $\hat{\eta}_i=\bigl(\hat{\eta}_i^1,\ldots,\hat{\eta}_i^{d+k}\bigr)$ and $\hat{\kappa}_i=\bigl(\hat{\kappa}_i^1,\ldots,\hat{\kappa}_i^{d+k}\bigr)$ defined as follows.

\par\bigskip
\noindent
	$\displaystyle
	\hat{\eta}^\ell_{i}=
	\left[\sum_{j=1}^{n_i}Z_{i\ell}(X_{ij})\hspace{-.2ex}+\hspace{-.2ex}\sum_{\stackrel{m,s: X_{ms}\leq X_{ij},}{\Delta_{ms}\eps_{ms}=1}}Z_{i\ell}(X_{ms})e^{\beta^TZ_i(X_{ms})}\hspace{-.2ex} \Lambda_n\{X_{ms}\}
	\frac{\mathcal G''\left(\int_0^{X_{ij}}e^{\beta_0^TZ_i(u)}d\Lambda_n(u)\right)}
	{\mathcal G'\left(\int_0^{X_{ij}}e^{\beta^TZ_i(u)}d\Lambda_n(u)\right)}\right]$
	\begin{eqnarray*}
		&&\hspace{11cm}\times\mathds{1}(\Delta_{ij}\eps_{ij}=2)\\[1.1ex]
		&&\hspace{1cm}-\sum_{j=1}^{n_i}\sum_{\stackrel{m,s: X_{ms}\leq X_{ij}\wedge\tau}{\Delta_{ms}\eps_{ms}=1}}Z_{i\ell}(X_{ms})e^{\beta^TZ_i(X_{ms})}\Lambda_n\{X_{ms}\}\,\mathcal G'\left(\int_0^{X_{ij}\wedge \tau} e^{\beta^TZ_i(u)}d\Lambda_n(u)\right)\\[1.1ex]
		&&-\,\left[\int_{D_i}^\tau
		\widetilde{w}_i^*(t)e^{\beta^TZ_i(t)}Z_{_{i\ell}}(t)\mathcal G'\left(\int_0^t e^{\beta^TZ_i(u)}d\Lambda_n(u)\right)d\Lambda_n(t)\right]\mathds{1}(\Delta_{ij}\eps_{ij}=2)
		\\[2.2ex]
		&&-\,\left[\int_{D_i}^\tau\hspace{-0.7ex}\widetilde{w}_i^*(t)e^{\beta^TZ_i(t)}
		\hspace{-0.2ex}\left(\int_0^t Z_{_{i\ell}}(u)e^{\beta^TZ_i(u)}d\Lambda_n(u)\right)\hspace{-0.2ex}
		\mathcal G''\left(\int_0^t e^{\beta^TZ_i(u)}d\Lambda_n(u)\right)d\Lambda_n(t) \right]\\[2ex]
		&&\hspace{11cm}\times\mathds{1}(\Delta_{ij}\eps_{ij}=2)
	\end{eqnarray*}

for \,$\ell\in\{1,\ldots,d\}$\, and

	\begin{eqnarray*}
		\hat{\eta}^\ell_{i}&=&\displaystyle\sum_{j=1}^{n_i}\left[\mathds{1}(X_\ell=X_{ij})\bigl[\Lambda_n\{X_\ell\}\bigr]^{-1}
		+\mathds{1}\bigl(X_\ell\leq (X_{ij}\wedge\tau))e^{\beta^TZ_i(X_\ell)}\phantom{\int_0^{X_{ij}}}\right.\\[0.8ex]
		&&\hspace{1.5cm}\displaystyle\left.\times
		\mathcal G''\biggl(\int_0^{X_{ij}}e^{\beta^TZ_i(u)}d\Lambda_n(u)\biggr)
		\Big/\mathcal G'\biggl(\int_0^{X_{ij}}e^{\beta^TZ_i(u)}d\Lambda_n(u)\biggr)\right]\mathds{1}(\Delta_{ij}\eps_{ij}=1)\\[1ex]
		&&-\,\sum_{j=1}^{n_i}\mathds{1}\bigl(X_\ell\leq(X_{ij}\wedge\tau)\bigr)e^{\beta^TZ_i(X_\ell)}\mathcal
		G'\biggl(\int_0^{X_\ell}e^{\beta^TZ_i(u)}d\Lambda_n(u)\biggr)\\[1.8ex]
		&&-\,\mathds{1}(D_i\leq X_\ell)\left[\widetilde{w}_i^*(X_\ell)e^{\beta^TZ_i(X_\ell)}\mathcal
		G'\biggl(\int_0^{X_\ell}e^{\beta^TZ_i(u)}d\Lambda_n(u)\biggr)\right]\mathds{1}(\Delta_{ij}\eps_{ij}=2)\\[1.5ex]
		&&-\,\left[e^{\beta^TZ_i(X_\ell)}\int_{D_i\wedge X_\ell}^\tau\widetilde{w}_i^*(t)
		e^{\beta^TZ_i(t)}\,\mathcal G''\biggl(\int_0^t e^{\beta^TZ_i(u)}d\Lambda_n(u)\biggr)d\Lambda_n(t)\right]\mathds{1}(\Delta_{ij}\eps_{ij}=2)
	\end{eqnarray*}
\\[0.5ex]
for  $\ell\in\{d+1,\ldots,d+k(n)\}$. To calculate $\hat{\kappa}_{i}^\ell$ we apply 

\begin{equation*}
\frac{\widehat{G}_c(t)}{\widehat{G}_c(X_{ij})}-\frac{G_c(t)}{G_c(X_{ij})}
	=-\frac{G_c(t)}{G_c(X_{ij})}\sum_{i=1}^n\int_{X_{ij}}^t\frac{1}{\sum_{k=1}^n\mathds{1}(X_k\hspace{-0.3ex}\geq\hspace{-0.3ex}u)}\hspace{-0.3ex}\times\hspace{-0.3ex} dM_i^c(u)+o_p(n^{-1/2}),
\end{equation*}
where $M_i^c(t)=\mathds{1} (X_i\leq t,\,\Delta_i=0)-\int_0^t\mathds{1}(X_i\geq u)dA^c(u)$ is 
the martingale associated with the censoring process and $A^c(t)$ is the cumulative hazard of the censoring distribution. From this we obtain the representation

\begin{equation*}
	\hat{\kappa}_{i}^\ell=\int_0^\infty
	q^\ell_n(u)\{\pi_n(u)\}^{-1}dM_i^c(u),\hee \mbox{with}
\end{equation*}

	\begin{eqnarray*} 
		q^\ell_n(u)&=&\frac{1}{n}\sum_{i=1}^n\sum_{j:\Delta_{ij}\eps_{ij}=2}\left[\int_0^\tau
		\widetilde{w}_i^*(t)\mathds{1}(X_{ij}\leq u\leq
		t)e^{\beta^TZ_i(t)}Z_{_{i\ell}}(t)\mathcal G'\left(\int_0^t e^{\beta^TZ_i(s)}d\Lambda_n(s)\right)d\Lambda_n(t)\right]\\[1ex]
		&&+\frac{1}{n}\sum_{i=1}^n\sum_{j:\Delta_{ij}\eps_{ij}=2}\left[
		\int_0^\tau\widetilde{w}_i^*(t)\mathds{1}(X_{ij}\leq u\leq t)e^{\beta^TZ_i(t)}\,
		\left(\int_0^t Z_{_{i\ell}}(s)e^{\beta^TZ_i(s)}d\Lambda_n(s)\right)\right.\\[1ex]
		&&\left.\hspace{2cm}\times\mathcal G''\left(\int_0^t
		e^{\beta^TZ_i(s)}d\Lambda_n(s)\right)d\Lambda_n(t)\right]\he
		\mbox{for} 
		\he\ell\in\{1,\ldots,d\},
		\\[2ex] q^\ell_n(u)&=&\frac{1}{n}\sum_{i=1}^n\sum_{j:\Delta_{ij}\eps_{ij}=2}\left[\widetilde{w}_i^*(X_\ell)\mathds{1}(X_{ij}\leq
	u\leq X_\ell)e^{\beta^TZ_i(X_\ell)}\,
	\mathcal G'\bigl(\int_0^{X_\ell}e^{\beta^TZ_i(s)}d\Lambda_n(s)\bigr)\right]\\[1ex]
		&&+\displaystyle\frac{1}{n}\sum_{i=1}^n\sum_{j:\Delta_{ij}\eps_{ij}=2}
		\left[e^{\beta^TZ_i(X_\ell)}\int_0^\tau\mathds{1}(X_\ell\leq t)\mathds{1}(X_{ij}\leq u\leq t)\widetilde{w}_i^*(t)
		e^{\beta^TZ_i(t)} \right.\\[1ex]
		&&\left.\hspace{3cm}\times \mathcal G''\bigl(\int_0^te^{\beta^TZ_i(s)}d\Lambda_n(s)\bigr)d\Lambda_n(t)\right]
	\end{eqnarray*}
\\[1ex]
for\, $\ell\in\{d+1,\ldots,d+k(n)\}$\, and\, $\pi_n(u)=n^{-1}\sum_{i=1}^n\sum_{j=1}^{n_i}\mathds{1}(X_{ij}\geq u)$.\\[2ex]

\subsection{Simulation Studies}\label{sec:suppsim}
We simulated data sequentially by drawing from the Gompertz distribution for both events, with a cumulative baseline rate $\Lambda_0^{\gamma_k,\gamma_\ell}(t)=\gamma_k\left\{1-\exp(-\gamma_\ell t)\right\}$. For the first recurrent event, $\varepsilon_{i1}\in\{1,2\}$ was 
generated from a Bernoulli distribution with parameter $F_1(\infty\mid Z)$. 
We generated the first recurrent event time from 
$F_1(t|Z)=1-\exp\bigl[-\mathcal G\{e^{\beta_1^TZ}\,\Lambda_0^{\gamma_1,\gamma_2}(t)\}\bigr]$.
The terminal event was generated from 
$F_2(t|Z)=1-\exp\bigl[-\mathcal G\{e^{\beta_2^TZ}\,\Lambda_0^{\gamma_3,\gamma_4}(t)\}\bigr]$, 
with $\gamma_3=\min\bigl(e^{-\beta_2^TZ}
\bigl[\{1-\rho\cdot\log F_1(\infty|Z)\}^{1/\rho}
-1\bigr], 0.3\bigr)$ for the Box--Cox transformation models and with $\gamma_3=\min\bigl(\left\{F_1(\infty|Z)^{-1/r}-1\right\}\cdot(r\cdot e^{\beta_2^TZ})^{-1}, 0.5\bigr)$ for the logarithmic transformation models.
For subsequent recurrences we generated $\eps_{ij}$ from the conditional probabilities $P(\eps_{ij}=1|T>t_{i_{j-1}},Z)=P(T>t_{i_{j-1}},\eps=1|Z)/P(T>t_{i_{j-1}}|Z)$. 
Subsequent recurrent event times were generated from $P(T\leq t_{i_j},\eps=1|T>t_{i_{j-1}},Z)=P(t_{i_{j-1}}<T\leq t_{ij},\eps=1|Z)/P(T>t_{i_{j-1}}|Z)$.

\begin{table}[!htbp]
	\centering
	\footnotesize
	\setlength{\tabcolsep}{6pt}
	\renewcommand{\arraystretch}{1.1}
	\caption{\small \textbf{Censoring distribution for each simulation scenario.} In all scenarios the study end is $\tau = 5$. The upper bound of the uniform distribution was chosen to yield approximately 15\% censoring in each scenario}
	\label{tab:sim_censoring}
	\begin{tabular}{llc}
		Model & Scenario & Censoring distribution \\[3pt]
		Box--Cox & $\rho = 0.2$ & $\min(C, \tau)$, $C \sim \mathrm{Uniform}(2, 20)$ \\   
		Box--Cox & $\rho = 0.5$ & $\min(C, \tau)$, $C \sim \mathrm{Uniform}(2, 20)$ \\
		Box--Cox & $\rho = 1$   & $\min(C, \tau)$, $C \sim \mathrm{Uniform}(2, 20)$ \\
		Box--Cox & $\rho = 2$   & $\min(C, \tau)$, $C \sim \mathrm{Uniform}(2, 20)$ \\
		Log     & $r = 0.5$    & $\min(C, \tau)$, $C \sim \mathrm{Uniform}(2, 21)$ \\
		Log     & $r = 1$      & $\min(C, \tau)$, $C \sim \mathrm{Uniform}(2, 20.5)$ \\
		Log      & $r = 2$      & $\min(C, \tau)$, $C \sim \mathrm{Uniform}(2, 21)$ \\   
	\end{tabular}
\end{table}


\begin{table}[!htbp]
\centering\footnotesize\setlength{\tabcolsep}{3.8pt}
\caption{Selection of the Box--Cox parameter by the TIC and the AIC.}
\label{tab:tic}
\begin{tabular}{cllcccccccc}
 & $n$ & & Mean & Median & SD & 95\% range & $\pm0.2$ & $\pm0.5$ & True rej. & GL rej.\\
\addlinespace[4pt]
\multirow{8}{*}{\rotatebox[origin=c]{90}{$\rho=0.2$}}
 & 50 & TIC & 0.26 & 0.1 & 0.28 & [0.1, 1.0] & 0.83 & 0.94 & 0.26 & 0.68\\
 & & AIC & 1.09 & 0.3 & 1.50 & [0.1, 4.0] & 0.58 & 0.71 & 0.26 & 0.42\\
\addlinespace[2pt]
 & 100 & TIC & 0.24 & 0.2 & 0.19 & [0.1, 0.7] & 0.88 & 0.97 & 0.21 & 0.84\\
 & & AIC & 0.42 & 0.3 & 0.60 & [0.1, 1.6] & 0.74 & 0.91 & 0.22 & 0.66\\
\addlinespace[2pt]
 & 200 & TIC & 0.22 & 0.2 & 0.13 & [0.1, 0.6] & 0.94 & 1.00 & 0.20 & 0.96\\
 & & AIC & 0.27 & 0.2 & 0.17 & [0.1, 0.7] & 0.88 & 0.98 & 0.24 & 0.93\\
\addlinespace[2pt]
 & 400 & TIC & 0.21 & 0.2 & 0.10 & [0.1, 0.4] & 0.97 & 1.00 & 0.21 & 1.00\\
 & & AIC & 0.23 & 0.2 & 0.11 & [0.1, 0.5] & 0.96 & 1.00 & 0.25 & 0.99\\
\addlinespace[8pt]
\multirow{8}{*}{\rotatebox[origin=c]{90}{$\rho=0.5$}}
 & 50 & TIC & 0.52 & 0.4 & 0.41 & [0.1, 1.5] & 0.59 & 0.93 & 0.23 & 0.53\\
 & & AIC & 1.25 & 0.7 & 1.35 & [0.1, 4.0] & 0.47 & 0.70 & 0.28 & 0.33\\
\addlinespace[2pt]
 & 100 & TIC & 0.51 & 0.5 & 0.26 & [0.2, 1.1] & 0.77 & 0.97 & 0.22 & 0.66\\
 & & AIC & 0.70 & 0.6 & 0.59 & [0.2, 2.3] & 0.68 & 0.88 & 0.25 & 0.50\\
\addlinespace[2pt]
 & 200 & TIC & 0.50 & 0.5 & 0.17 & [0.3, 0.9] & 0.91 & 0.99 & 0.23 & 0.85\\
 & & AIC & 0.56 & 0.5 & 0.19 & [0.3, 1.0] & 0.86 & 0.97 & 0.27 & 0.73\\
\addlinespace[2pt]
 & 400 & TIC & 0.50 & 0.5 & 0.11 & [0.3, 0.7] & 0.97 & 1.00 & 0.25 & 0.97\\
 & & AIC & 0.52 & 0.5 & 0.12 & [0.3, 0.8] & 0.96 & 1.00 & 0.29 & 0.94\\
\addlinespace[8pt]
\multirow{8}{*}{\rotatebox[origin=c]{90}{$\rho=1$}}
 & 50 & TIC & 1.02 & 0.9 & 0.68 & [0.2, 3.0] & 0.37 & 0.70 & 0.24 & --\\
 & & AIC & 1.98 & 1.3 & 1.41 & [0.4, 4.0] & 0.26 & 0.51 & 0.22 & --\\
\addlinespace[2pt]
 & 100 & TIC & 1.01 & 0.9 & 0.48 & [0.4, 2.2] & 0.48 & 0.86 & 0.22 & --\\
 & & AIC & 1.37 & 1.1 & 0.88 & [0.5, 4.0] & 0.42 & 0.74 & 0.23 & --\\
\addlinespace[2pt]
 & 200 & TIC & 0.99 & 0.9 & 0.30 & [0.6, 1.7] & 0.64 & 0.95 & 0.20 & --\\
 & & AIC & 1.11 & 1.0 & 0.40 & [0.6, 2.1] & 0.59 & 0.89 & 0.18 & --\\
\addlinespace[2pt]
 & 400 & TIC & 0.97 & 1.0 & 0.18 & [0.7, 1.4] & 0.84 & 0.99 & 0.17 & --\\
 & & AIC & 1.01 & 1.0 & 0.20 & [0.7, 1.5] & 0.83 & 0.99 & 0.15 & --\\
\addlinespace[8pt]
\multirow{8}{*}{\rotatebox[origin=c]{90}{$\rho=2$}}
 & 50 & TIC & 1.91 & 1.6 & 0.99 & [0.7, 4.0] & 0.16 & 0.36 & 0.18 & 0.41\\
 & & AIC & 2.93 & 4.0 & 1.22 & [0.8, 4.0] & 0.10 & 0.23 & 0.17 & 0.61\\
\addlinespace[2pt]
 & 100 & TIC & 1.92 & 1.7 & 0.79 & [0.9, 4.0] & 0.23 & 0.50 & 0.19 & 0.62\\
 & & AIC & 2.59 & 2.3 & 1.08 & [1.0, 4.0] & 0.16 & 0.40 & 0.18 & 0.75\\
\addlinespace[2pt]
 & 200 & TIC & 1.96 & 1.8 & 0.60 & [1.1, 3.6] & 0.32 & 0.67 & 0.19 & 0.88\\
 & & AIC & 2.30 & 2.1 & 0.79 & [1.2, 4.0] & 0.28 & 0.57 & 0.17 & 0.92\\
\addlinespace[2pt]
 & 400 & TIC & 1.99 & 1.9 & 0.47 & [1.3, 3.2] & 0.43 & 0.82 & 0.18 & 0.99\\
 & & AIC & 2.15 & 2.0 & 0.57 & [1.4, 3.8] & 0.42 & 0.77 & 0.16 & 0.99\\
\end{tabular}
\par\smallskip
{\raggedright $\rho$, true Box--Cox parameter ($\rho=1$: Ghosh--Lin model); mean, median, standard
deviation (SD) and 95\% range of the selected $\hat\rho$; $\pm0.2$,
$\pm0.5$: proportion of replicates with $|\hat\rho-\rho|\le0.2$ and $\le0.5$; True rej.: proportion of
replicates in which the true model had a criterion value at least 2 units above the best candidate; GL rej.:
the same for the Ghosh--Lin model when misspecified. $\hat\rho$ is selected on a grid with step 0.1,
$0.1\le\hat\rho\le4$.\par}
\end{table}

Data were generated from Box--Cox transformation models with $\rho\in\{0.2,0.5,1,2\}$, with
1000 replicates per setting and the remaining settings as in Table~1 of the main paper
(for $\rho=0.2$, those of $\rho=0.5$). In each replicate, Box--Cox models with
$\rho\in\{0.1,0.2,\ldots,4.0\}$ and logarithmic models with $r\in\{0.1,0.2,\ldots,3.0\}$ were
fitted, and the models with the smallest TIC and AIC were selected (Table~\ref{tab:tic}).

The TIC selected the transformation more accurately than the AIC, in particular for small
sample sizes, where the AIC frequently selected the upper end of the grid. The accuracy of
both criteria increased with the sample size; for $n=400$, the TIC-selected $\hat\rho$ satisfied $|\hat\rho-\rho|\le0.5$ in 82\% to 100\% of the replicates, and the misspecified
Ghosh--Lin model was rejected in 97\% to 100\% of the replicates. The standard deviation of the selected parameter was approximately proportional to $\rho$,
0.11, 0.18 and 0.47 for $\rho=0.5$, 1 and 2 at $n=400$, so that the larger absolute
deviations for $\rho=2$ reflect the scale of $\rho$.



\begin{table}[!htbp]
\centering\footnotesize\setlength{\tabcolsep}{5pt}
\caption{Coverage of 95\% confidence intervals for the marginal mean: true model / AIC-selected / TIC-selected / fixed Ghosh--Lin model.}
\label{tab:tic_cov}
\begin{tabular}{clcccc}
 & & $n=50$ & $n=100$ & $n=200$ & $n=400$\\
\addlinespace[4pt]
\multirow{6}{*}{\rotatebox[origin=c]{90}{$\rho=0.2$}}
 & $m_0(\tau/4)$ & 0.92/0.91/0.91/0.93 & 0.93/0.93/0.94/0.94 & 0.95/0.95/0.94/0.95 & 0.95/0.94/0.94/0.95\\
 & $m_0(\tau/2)$ & 0.93/0.92/0.93/0.93 & 0.94/0.93/0.94/0.93 & 0.95/0.95/0.95/0.93 & 0.95/0.94/0.95/0.93\\
 & $m_0(\tau)$ & 0.93/0.93/0.93/0.91 & 0.94/0.93/0.92/0.91 & 0.94/0.93/0.93/0.91 & 0.95/0.94/0.94/0.89\\
 & $m_1(\tau/4)$ & 0.92/0.90/0.90/0.86 & 0.94/0.92/0.93/0.83 & 0.95/0.95/0.95/0.80 & 0.96/0.95/0.96/0.65\\
 & $m_1(\tau/2)$ & 0.93/0.91/0.89/0.88 & 0.94/0.93/0.93/0.87 & 0.95/0.95/0.94/0.85 & 0.97/0.96/0.96/0.75\\
 & $m_1(\tau)$ & 0.92/0.91/0.88/0.91 & 0.94/0.93/0.92/0.90 & 0.94/0.94/0.94/0.90 & 0.95/0.94/0.95/0.86\\
\addlinespace[8pt]
\multirow{6}{*}{\rotatebox[origin=c]{90}{$\rho=0.5$}}
 & $m_0(\tau/4)$ & 0.92/0.91/0.91/0.94 & 0.94/0.93/0.93/0.94 & 0.96/0.94/0.94/0.95 & 0.94/0.93/0.93/0.91\\
 & $m_0(\tau/2)$ & 0.93/0.92/0.92/0.93 & 0.94/0.94/0.95/0.95 & 0.96/0.95/0.96/0.96 & 0.95/0.95/0.95/0.95\\
 & $m_0(\tau)$ & 0.94/0.92/0.92/0.92 & 0.95/0.94/0.94/0.95 & 0.96/0.96/0.95/0.95 & 0.94/0.94/0.94/0.94\\
 & $m_1(\tau/4)$ & 0.91/0.88/0.90/0.87 & 0.94/0.91/0.93/0.87 & 0.95/0.94/0.94/0.86 & 0.95/0.93/0.93/0.78\\
 & $m_1(\tau/2)$ & 0.92/0.89/0.91/0.90 & 0.93/0.92/0.93/0.90 & 0.95/0.94/0.94/0.89 & 0.95/0.93/0.93/0.83\\
 & $m_1(\tau)$ & 0.91/0.89/0.89/0.90 & 0.94/0.92/0.93/0.92 & 0.94/0.93/0.94/0.92 & 0.94/0.92/0.93/0.89\\
\addlinespace[8pt]
\multirow{6}{*}{\rotatebox[origin=c]{90}{$\rho=1$}}
 & $m_0(\tau/4)$ & 0.94/0.89/0.89/-- & 0.94/0.90/0.90/-- & 0.94/0.90/0.90/-- & 0.95/0.93/0.91/--\\
 & $m_0(\tau/2)$ & 0.94/0.93/0.93/-- & 0.94/0.93/0.92/-- & 0.94/0.93/0.93/-- & 0.96/0.95/0.95/--\\
 & $m_0(\tau)$ & 0.94/0.93/0.93/-- & 0.94/0.95/0.94/-- & 0.95/0.94/0.94/-- & 0.95/0.94/0.94/--\\
 & $m_1(\tau/4)$ & 0.94/0.92/0.93/-- & 0.94/0.93/0.94/-- & 0.93/0.92/0.94/-- & 0.94/0.93/0.93/--\\
 & $m_1(\tau/2)$ & 0.92/0.90/0.92/-- & 0.93/0.92/0.92/-- & 0.95/0.93/0.93/-- & 0.94/0.93/0.93/--\\
 & $m_1(\tau)$ & 0.92/0.88/0.89/-- & 0.93/0.91/0.91/-- & 0.95/0.91/0.91/-- & 0.94/0.91/0.90/--\\
\addlinespace[8pt]
\multirow{6}{*}{\rotatebox[origin=c]{90}{$\rho=2$}}
 & $m_0(\tau/4)$ & 0.93/0.89/0.87/0.80 & 0.95/0.91/0.90/0.76 & 0.96/0.90/0.91/0.62 & 0.96/0.90/0.90/0.41\\
 & $m_0(\tau/2)$ & 0.92/0.91/0.91/0.89 & 0.95/0.93/0.93/0.91 & 0.96/0.93/0.94/0.89 & 0.95/0.92/0.92/0.83\\
 & $m_0(\tau)$ & 0.94/0.93/0.94/0.93 & 0.94/0.94/0.94/0.95 & 0.95/0.94/0.95/0.94 & 0.94/0.95/0.95/0.94\\
 & $m_1(\tau/4)$ & 0.93/0.91/0.92/0.93 & 0.94/0.93/0.93/0.92 & 0.95/0.94/0.94/0.91 & 0.95/0.94/0.94/0.84\\
 & $m_1(\tau/2)$ & 0.91/0.90/0.90/0.91 & 0.94/0.94/0.93/0.89 & 0.95/0.93/0.93/0.82 & 0.95/0.94/0.94/0.68\\
 & $m_1(\tau)$ & 0.92/0.89/0.89/0.87 & 0.94/0.92/0.92/0.80 & 0.94/0.93/0.92/0.67 & 0.95/0.92/0.91/0.45\\
\end{tabular}
\par\smallskip
{\raggedright $m_0(t)$ and $m_1(t)$: marginal mean $\mu(t\mid Z)=\mathcal G\{\exp(\beta^TZ)\Lambda_0(t)\}$
for $Z=(0,0)$ and $Z=(1,0)$, at $t=\tau/4$, $\tau/2$ and $\tau$. Confidence intervals based on the
sandwich estimator for the selected model, without adjustment for the selection. For $\rho=1$ the
fixed Ghosh--Lin model is the true model.\par}
\end{table}

Table~\ref{tab:tic_cov} shows the coverage of 95\% confidence intervals for the marginal
mean. After selection by the Takeuchi criterion, coverage was close to that under the true model, between 0.90 and 0.96 for $n=400$, although the intervals do not account for the variability of the selection. The fixed Ghosh--Lin model, in contrast, undercovered under misspecification, and
its coverage decreased with increasing sample size, to 0.41 for $\rho=2$, 0.65 for $\rho=0.2$
and 0.78 for $\rho=0.5$ at $n=400$, reflecting a bias that does not vanish as the sample size
increases. The Akaike criterion gave similar results to the Takeuchi criterion.

\subsection{Statcope Data Analysis}
\label{supp:supp_statdata}
We provide additional material for the STATCOPE trial interim futility analysis.\\

\begin{table}[!htbp]
	\centering
	\footnotesize
	\caption{\footnotesize Exacerbation counts for the STATCOPE trial}
	\label{tab:statcope_exacerbations}
	\begin{tabular}{lrrrr}
		& {\bf All} & {\bf GOLD 2} & {\bf GOLD 3} & {\bf GOLD 4} \\[3pt]
		Subjects & 870 & 286 & 297 & 287 \\
		Deaths, no. (\%) & 58 (6.7) & 5 (1.7) & 16 (5.4) & 37 (12.9) \\[0.5ex]
		Total exacerbations & 1950 & 486 & 643 & 821 \\
		\quad Severity 1 (mild) & 1253 & 319 & 426 & 508 \\
		\quad Severity 2 (moderate) & 261 & 86 & 89 & 86 \\
		\quad Severity 3 (severe) & 406 & 79 & 121 & 206 \\
		\quad Severity 4 (very severe) & 30 & 2 & 7 & 21 \\[1ex]
	\end{tabular}
	
	\begin{minipage}{\linewidth}
		\footnotesize
		Severity 1: mild exacerbations requiring home management. 
		Severity 2: moderate exacerbations requiring an emergency department visit. 
		Severity 3: severe exacerbations requiring hospital admission. 
		Severity 4: very severe exacerbations requiring intubation and mechanical ventilation.
	\end{minipage}
\end{table}

\begin{table}[!htbp]
	\centering
	\footnotesize
	\label{tab:exacerbation_dist}
	\setlength{\tabcolsep}{4pt}
	\renewcommand{\arraystretch}{1.1}
	\caption{\footnotesize Distribution of exacerbation counts per subject, severity 3--4}
	\begin{tabular}{lrrrrrrrrrrr}
		{\bf Exacerbations} & 0 & 1 & 2 & 3 & 4 & 5 & 6 & 7 & 9 & 10 & Sum \\[3pt]
		GOLD 2--4 & 639 & 138 & 45 & 28 & 5 & 5 & 3 & 2 & 3 & 2 & 436 \\
		GOLD 4    & 172 & 66  & 22 & 14 & 4 & 3 & 2 & 2 & 2 & -- & 227 \\
	\end{tabular}
\end{table}

\begin{table}[!htbp]
\centering
\footnotesize
\label{tab:predict1100}
\caption{\footnotesize STATCOPE data: Predicted Marginal Mean for Different Transformation Models}
\begin{tabular}{lrrr}
 & {\bf Placebo} & {\bf Simvastatin} & {\bf Reduction}\\[0.5ex]
Box--Cox, $\rho = 1$ & 2.88 & 2.54 & 11.8\%\\
Box--Cox, $\rho = 0.18$ & 2.45 & 2.35 & 4.3\%\\
Logarithmic, $r = 0.64$ & 2.45 & 2.35 & 4.3\%\\
Logarithmic, $r = 1$ & 2.25 & 2.20 & 2.3\%\\[0.7ex]
\end{tabular}

\begin{minipage}{\linewidth}
\footnotesize
    Predicted marginal mean of severe exacerbations after 1100 days for a former smoker aged $< 70$ with stage 4 disease, 40 pack-years (median), prior hospitalization for chronic obstructive pulmonary disease and supplemental oxygen at baseline, under placebo and simvastatin for different transformation models.
\end{minipage}
\end{table}

\subsection{Readmission Data Analysis}
\label{supp:readmission}
The readmission data (Gonz\'alez et al., 2005) comprise 403 patients with colorectal cancer that were followed up for recurrent hospitalizations until death, censoring or endpoint of study at 2176 days. Time-fixed covariates reported were chemotherapy as a treatment, sex, Dukes stage and Charlson comorbidity index at baseline (Table 11). The study reported 458 readmissions and 109 deaths (27.0\%) (Table 10). The optimal transformation models were the Box--Cox transformation model with $\rho = 0.37$ and the logarithmic transformation model with $r = 0.39$, selected by the Takeuchi information criterion.

The estimated regression coefficients from the Andersen--Gill model and from the negative binomial regression differ markedly from the pseudo-likelihood approach (Table 12): the coefficient for Dukes stage D is 0.551 under the identity transformation, but 1.314 under the Andersen--Gill model and 1.494 under the negative binomial model. The estimated coefficient for chemotherapy as treatment is $-0.497$ under the identity transformation, compared with $-0.242$ and $-0.159$ under the Andersen--Gill model and the negative binomial model. Both covariates are strongly associated with death, with Cox hazard ratios of 29.6 for Dukes stage D and 2.16 for chemotherapy. The Andersen--Gill model and negative binomial regression estimate the readmission rate conditional on survival, whereas the pseudo-likelihood approach estimates the marginal mean, which is lower for patients at higher risk of death because patients who die can no longer be readmitted.

For a male patient with Dukes stage D and Charlson index $\ge 3$ without chemotherapy, the pseudo-likelihood approach predicts 3.07 ($\rho = 1$) and 2.79 ($\rho = 0.37$) readmissions at 3 years, whereas the Zeng--Lin model and negative binomial regression predict 5.77 and 8.00 (Fig. 5). The predicted reduction under chemotherapy ranges from 39.2\% ($\rho = 1$) to 20.7\% ($r = 1$) (Table 13).
\begin{table}[!htbp]
\centering
\footnotesize
\caption{\footnotesize Readmission data: distribution of readmission counts per patient.}
\begin{tabular}{lrrrrrrrrrrrrrr}
Readmissions & 0 & 1 & 2 & 3 & 4 & 5 & 6 & 8 & 9 & 10 & 11 & 16 & 22 & Sum\\[0.5ex]
Patients & 199 & 105 & 45 & 21 & 15 & 8 & 4 & 1 & 1 & 1 & 1 & 1 & 1 & 458\\
\end{tabular}
\end{table}

\begin{table}[!htbp]
\centering
\footnotesize
\label{tab:readm7}
\caption{\footnotesize Readmission data: patients, deaths and readmissions by Dukes stage.}
\begin{tabular}{lrrrr}
 &{\bf All} & {\bf Dukes A--B} & {\bf Dukes C} & {\bf Dukes D}\\[0.5ex]
Patients & 403 & 180 & 148 & 75\\
Deaths, no. (\%) & 109 (27.0) & 15 (8.3) & 36 (24.3) & 58 (77.3)\\
Readmissions & 458 & 144 & 183 & 131\\
Chemotherapy, no. (\%) & 217 (53.8) & 129 (71.7) & 44 (29.7) & 44 (58.7)\\[0.7ex]
\end{tabular}
\end{table}

\begin{table}[!htbp]
\centering
\footnotesize
\setlength{\tabcolsep}{2.2pt}
\caption{Readmission data: parameter estimates and standard errors for different regression models}

\label{tab:readm_models}
\resizebox{\textwidth}{!}{%
\begin{tabular}{lrrrrrrr}
 & \multicolumn{2}{c}{\bf Pseudo-likelihood: logarithmic} & \multicolumn{2}{c}{\bf Pseudo-likelihood: Box--Cox} & {\bf Ghosh--Lin} & {\bf Andersen--Gill} & {\bf Neg.\ binomial}\\[0.5ex]
 & $r = 1$ & $r = 0.39$ & $\rho = 0.37$ & $\rho = 1$ &  &  & \\
$\Delta$TIC & 5.36 & 0.00 & 0.72 & 3.67 &  &  & \\
chemo & $-$0.552 (0.239) & $-$0.543 (0.195) & $-$0.573 (0.204) & {\bf$-$0.497 (0.164)} & {\bf$-$0.497 (0.164)} & {\bf$-$0.242 (0.156)} & {\bf$-$0.159 (0.159)}\\
female & $-$0.731 (0.234) & $-$0.624 (0.202) & $-$0.647 (0.211) & $-$0.503 (0.171) & $-$0.503 (0.171) & $-$0.534 (0.173) & $-$0.627 (0.163)\\
Dukes C & 0.266 (0.252) & 0.188 (0.208) & 0.206 (0.218) & 0.159 (0.179) & 0.160 (0.179) & 0.346 (0.185) & 0.338 (0.185)\\
Dukes D & 0.918 (0.329) & 0.713 (0.264) & 0.742 (0.270) & 0.551 (0.203) & 0.552 (0.203) & 1.314 (0.214) & 1.494 (0.234)\\
Charlson1-2 & 1.372 (0.604) & 1.000 (0.375) & 1.049 (0.380) & 0.751 (0.247) & 0.752 (0.247) & 0.787 (0.266) & 1.025 (0.416)\\
Charlson$\ge 3$ & 0.708 (0.300) & 0.589 (0.240) & 0.616 (0.250) & 0.474 (0.186) & 0.475 (0.187) & 0.521 (0.196) & 0.494 (0.224)\\[0.5ex]
\end{tabular}}

	\vspace{4pt}
\begin{minipage}{\linewidth}
	\footnotesize
	Columns 2--5: Estimated parameters (standard errors) from the pseudo-likelihood approach under  the logarithmic transformations with $r = 1$, transformation models selected by the TIC (logarithmic with $r = 0.39$, Box--Cox with $\rho = 0.37$), and the identity link. Columns 6--8: Ghosh--Lin model, Andersen--Gill model, and negative binomial model. Covariates at baseline: chemo (chemotherapy), female (sex), Dukes stage C and D (reference A--B), Charlson index 1--2 and $\ge 3$ (reference 0).
\end{minipage}
\end{table}

\begin{figure}[!htbp]
	\centering
	\includegraphics[scale=0.8]{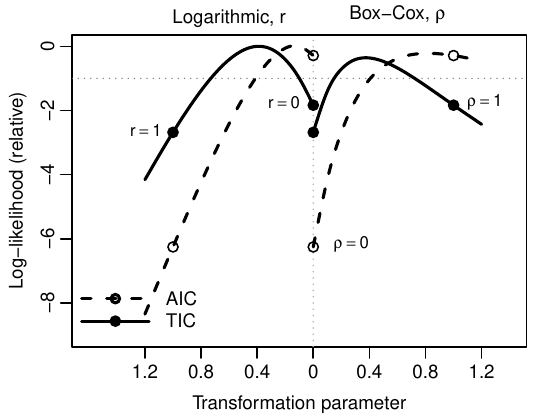}
	\caption{\footnotesize Readmission data: model selection using the Takeuchi and Akaike information criteria (TIC, AIC). Profile log-pseudo-likelihood curves for the logarithmic transformation models (left side) and Box--Cox transformation models (right side).}
	\label{fig:readm_loglik}
\end{figure}

\begin{figure}[!htbp]
	\centering
	\includegraphics[scale=0.6]{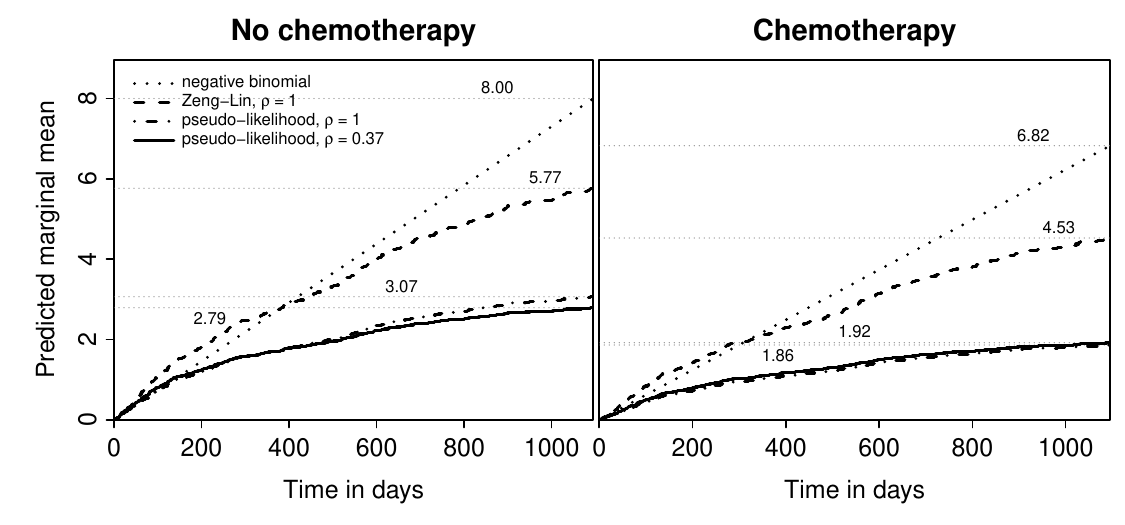}
	\caption{\footnotesize Readmission data: predicted marginal mean of readmissions for a male patient with Dukes stage D and Charlson index $\ge 3$, without (left panel) and with chemotherapy (right panel), from negative binomial regression, the Zeng--Lin model ($\rho = 1$) and the pseudo-likelihood approach under the identity transformation ($\rho = 1$) and the Box--Cox transformation selected by the Takeuchi information criterion ($\rho=0.37$). Dotted lines mark the values at 3 years.}
	\label{fig:readm_predict}
\end{figure}

\begin{table}[!htbp]
	\centering
	\footnotesize
	\caption{\footnotesize Readmission data: Individualized Prediction for the Marginal Mean}
	\label{tab:readm_predict}
	\begin{tabular}{lrrr}
		& No chemotherapy & Chemotherapy & Reduction \\[3pt]
		Box--Cox, $\rho = 1$       & 3.07 & 1.86 & 39.2\% \\
		Box--Cox, $\rho = 0.37$    & 2.79 & 1.92 & 31.1\% \\
		Logarithmic, $r = 0.39$    & 2.77 & 1.94 & 30.0\% \\
		Logarithmic, $r = 1$       & 2.33 & 1.85 & 20.7\% \\
	\end{tabular}
	\par\vspace{1mm}
	\parbox{\textwidth}{\footnotesize Predicted marginal mean of readmissions after 1095 days for a male patient with Dukes stage D and Charlson index $\ge 3$, without and with chemotherapy, for different transformation models.}
\end{table}

%
\begin{table}[!htbp]
	\centering
	\footnotesize
	\renewcommand{\arraystretch}{1.0}
	\caption{\footnotesize List of symbols}
	\label{tab:symbols}
	\begin{tabular}{p{4.3cm}p{8.7cm}p{1.4cm}}
		Symbol & Description & Section \\[3pt]
		\multicolumn{3}{l}{\textit{Rates, parameters and estimators}}\\[1pt]
		$\lambda\{t\mid Z(t)\}$ & Marginal rate function given the covariate history & 2 \\
		$\Lambda\{t\mid Z(t)\}$ & Cumulative marginal rate given the covariate history, model (1) & 2 \\
		$\Lambda(t)$, $\lambda(t)=d\Lambda(t)/dt$ & Cumulative baseline marginal rate and baseline marginal rate & 2 \\
		$\Lambda\{t\}=\Lambda(t)-\Lambda(t-)$ & Jump size of $\Lambda$ at $t$ & 3.3 \\
		$\lambda^*(t\mid Z)$ & Cause-specific rate conditional on survival & 2 \\
		$\mathcal G$ & Transformation (link) function & 2 \\
		$\rho$, $r$ & Parameters of the Box--Cox and logarithmic transformation models & 2 \\
		$\beta$, $\theta=(\beta,\Lambda)$ & Regression parameter and full model parameter & 2 \\
		$\beta_0$, $\Lambda_0$, $\theta_0=(\beta_0,\Lambda_0)$ & True parameter values & 4.3 \\
		$\Theta$ & Parameter space & A.2 \\
		$\widehat\beta_n$, $\widehat\Lambda_n$, $\widehat\theta_n$ & Pseudo-likelihood estimators of $\beta_0$, $\Lambda_0$ and $\theta_0$ & 4.3 \\
		$\widehat\Lambda(t)$ & Nelson--Aalen-type estimator of the cumulative marginal rate without covariates & 3.1 \\
		$\Lambda_n\{X_{ij}\}$ & Jump sizes of the step function approximating $\Lambda$ & B.1 \\
		$\widetilde\Lambda_n(t)=n^{-1}\sum_{i}N_i(t)$ & Comparison function in the existence proof & A.2 \\
		$\widetilde\Lambda^c_n(t)$ & Approximation of $\Lambda_0$ for recurrent and cure events & A.2 \\
		$\widetilde\Lambda\{\widetilde T_i\}$ & Jump sizes at the ordered recurrent event times $\widetilde T_1<\ldots<\widetilde T_k$ & 4.3 \\
		$\beta_*$, $\Lambda_*$, $\theta_*$ & Limits of convergent subsequences & A.2 \\
		$\Lambda_\theta(t\mid Z)$, $\lambda_\theta(t\mid Z)$, $S_\theta(t\mid Z)$ & Cumulative marginal rate, marginal rate and $\exp\{-\Lambda_\theta(t\mid Z)\}$ under $\theta$ & A.2 \\
		$\Lambda_0^{\gamma_k,\gamma_\ell}(t)$ & Gompertz-type cumulative baselines used for data generation & 5 \\
		$\ell(\beta,\Lambda)$, $\ell_n$ & Log-pseudo-likelihood function and its empirical version & 4.3 \\
		$\mathcal I_n$, $\widehat\Sigma_n$ & Observed Fisher information and estimated variance of the score & 4.3 \\
		$\hat\eta_i$, $\hat\kappa_i$ & Components of the iid decomposition of the score & B.3 \\
		TIC & Takeuchi information criterion & 4.3 \\[4pt]
		\multicolumn{3}{l}{\textit{Data, processes, weights and operators}}\\[1pt]
		$T_{ij}$, $\varepsilon_{ij}\in\{1,2\}$ & Event times and event type (recurrence or terminal event) & 3 \\
		$D$, $C$, $\tau$ & Terminal event time, censoring time and study endpoint & 3 \\
		$X_{ij}=T_{ij}\wedge C_i\wedge D_i$, $\Delta_{ij}$ & Observed times and indicators of an observed recurrence & 3.2 \\
		$N^*(t)$, $N_i^*(t)$ & Processes counting the recurrent events without censoring & 3 \\
		$N(t)$, $N_i(t)$ & Processes counting the observed recurrent events under censoring & 3.2 \\
		$M_i(t)$ & Mean-zero process from the decomposition of $dN_i$ & 3.1 \\
		$M_i^c(t)$, $A^c(t)$ & Censoring martingale and cumulative hazard of the censoring distribution & B.3 \\
		$Y_i(t)$, $Y(t)$ & At-risk indicator and size of the risk set & 3.1 \\
		$\mathcal R(t)$, $\mathcal R_p(t)$ & Risk set and pseudo-risk set & 3.1 \\
		$S(t)=\exp\{-\Lambda(t)\}$ & Probability of no recurrence until $t$ under the working assumption & 3.1 \\
		$S_D(t\mid Z)$ & Survival function of the terminal event & 2 \\
		$G_c$, $\widehat G_c$ & Censoring distribution and its Kaplan--Meier estimator & 3.3 \\
		$w_i(t)$, $w_i^*(t)$ & Inverse probability of censoring weight and simplified weight function & 3.3 \\
		$\widetilde w$, $\widehat w_n$ & Deterministic and estimated weight functions & A.3 \\
		$\Phi_n^c$, $\Phi_n^\omega$ & Denominators in the equations for the jump sizes, for cure and absorbing events & A.2 \\
		$\Phi_*^c$, $\Phi_*^\omega$, $\eta(s,\beta_0,\Lambda_0)$ & Limits of $\Phi_n^c$ and $\Phi_n^\omega$, and limiting numerator & A.2 \\
		$\mathcal K_c$, $\mathcal K_{\omega^*}$ & Kullback--Leibler information for cure and absorbing events & A.2 \\
		$\Psi_n$, $\Psi$, $\psi$ & Empirical and limiting score operators and score function & A.3 \\
		$\mathcal Z_1$, $\mathcal Z_2$ & Tight limits in the weak convergence of the score & A.3 \\
		$\mathcal H$, $h=(h_1,h_2)$ & Index set and its elements for linear functionals & 4.3 \\
		$M$ & Bound on $|\beta^TZ(t)|$ under condition M2) & A.1 \\
	\end{tabular}
\end{table}

\end{document}